# Global Inequalities in Clinical Trials Participation

Wen Lou[1,2,3], Adrián A. Díaz-Faes[4*], Jiangen He[5], Zhihao Liu[2], Vincent Larivière[6,7,8,9*]

1 School of Information and Communication, Nankai University, Tianjin, PR China
2 Department of Information Management, School of Economics and Management, East China Normal University, Shanghai, PR China
3 Key Laboratory of Advanced Theory and Application in Statistics and Data Science-MOE (East China Normal University), Shanghai, PR China
4 INGENIO, CSIC-Universitat Politècnica de València, Valencia, Spain
5 School of Information Sciences, The University of Tennessee, Knoxville, TN, USA
6 École de bibliothéconomie et des sciences de l'information, Université de Montréal, QC, Canada
7 Consortium Érudit, Montréal, Québec, Canada
8 Department of Science and Innovation-National Research Foundation Centre of Excellence in Scientometrics and Science, Technology and Innovation Policy, Stellenbosch University, Stellenbosch, Western Cape, South Africa
9 Observatoire des sciences et des technologies, Centre interuniversitaire de recherche sur la science et la technologie, Université du Québec à Montréal, Montréal, QC, Canada

## Abstract

Clinical trials are fundamental to the production of medical evidence and determine who gains access to experimental therapies. Although prior work has long documented inequalities in global clinical trial participation, a systematic quantification of the relative contributions of country-level factors and disease burden remains absent. This paper analyzes inequality in participation across 78,117 randomized controlled trials (RCTs) spanning 16 major disease categories from 2000 to 2024. Linking 185.6 million RCTs participants to country-level disease burden, the paper shows that inequalities in RCTs participation are predominantly explained by country-level factors rather than disease burden. Country-level factors account for 88.9% of the variation in participation, whereas disease and temporal effects contribute marginally. Removing entire disease categories, including those traditionally underfunded, has little effect on overall inequality. Instead, participation is highly concentrated geographically, with a small group of Global North countries enrolling a disproportionate share of participants across nearly all diseases. These patterns persist despite decades of disease-targeted funding and increasing alignment between research efforts and disease burden. Findings also show that without parallel horizontal investments in research capacity, health infrastructure, and governance, even well-funded disease programs are insufficient to reduce inequality in clinical trial participation.

## Introduction

Global health investments have grown significantly over the last decades, despite recent reductions in U.S. global health funding [1]. Health spending in low-income countries grew about 5% annually from 1995 to 2026, increasing from $51 to $153 per capita [2,3] while targeted funding initiatives (e.g., PEPFAR, The Global Fund) demonstrated measurable impact on specific diseases [4]. Yet profound inequalities persist across multiple dimensions (socio-economic, environmental, and healthcare access) of global health [5]. The global health community has responded to such inequalities primarily through vertical approaches centred on specific diseases. This means directing funding toward neglected and high-burden diseases and supporting targeted interventions such as vaccines, treatments, and public health campaigns [6]. International health aid increased by 30% from 2009 to 2024 [7] and HIV/AIDS research grew considerably following major funding commitments [3]. Yet, further progress is constrained by two factors: persistent structural inequalities in global health research, and recent stagnation of international support for global health [8,9].

Recent evidence shows an increasing alignment between research efforts and burden of disease. However, this trend seems to be driven more by changes in disease burden than by actual shifts in research direction [10], since switching costs are high [11]. Similarly, most drug innovation efforts concentrate on diseases prevalent in high-income countries and with large market size [12,13], although low- and middle-income countries account for 92% of the global disease burden [14]. Global health research capacity remains concentrated in the Global North, with less than 10% of health publications authored by researchers affiliated with institutions from the Global South [15]. As for health spending per capita, high-income countries in 2016 invested 130 times more than low-income countries, and is projected to persist at 126 times through 2050 [16]. Development assistance for health has grown only modestly, at 1.2% annually since 2010, and has stagnated around $39 billion for a decade, except for the peak during COVID-19 [6]. These numbers raise the question of whether participation inequality reflects the specific diseases being studied or deeper structural features of countries, a distinction with direct implications for epistemic justice in global health.

From a systemic perspective, vertical approaches based on disease-targeted interventions, while addressing immediate needs, may inadvertently perpetuate structural inequalities [17]. Interventions centered on specific diseases can create parallel bureaucracies, lead to inefficient facility utilization, and leave gaps in care, especially for patients with multiple co-morbidities [18]. Often funded by international and supranational organizations, vertical approaches tend to divert skilled local health personnel from general health services, creating internal brain drain and jeopardizing access to local health services [19]. Such competition for funding and recognition often orients researchers toward these international initiatives rather than strengthening national health systems [20]. These concerns do not negate the value of vertical programs; however, they show why such programs alone are unlikely to reduce participation inequality. Moreover, targeted funding may influence what is studied, but has little capacity to change where clinical trials are conducted and participants enrolled.

This paper examines global health inequalities through the lens of participation in terms of research production (who develops and performs research) and research participants[1]. To do so, we rely on randomized controlled trials (RCTs), which occupy a unique position in health research. RCTs are the gold standard in medicine and the pathway through which new therapeutics gain market access [21]. This dual role creates two distinct but related benefits. First, RCTs provide participating patients direct access to experimental treatments, advanced monitoring, and high-quality care, often representing the most sophisticated medical attention available. In many settings, though not exclusively, in low- and middle-income countries, participation in RCTs can provide access to investigational therapies that are otherwise financially or logistically inaccessible [22]. Second, RCTs generate the evidence that determines which treatments become standard practice, thereby shaping healthcare delivery across all populations and settings [23].

[1] Note that in 2024, Declaration of Helsinki has changed the term "Subject" to "Participant". World Medical Association. (2024). Declaration of Helsinki: Ethical principles for medical research involving human participants. https://www.wma.net/policies-post/wma-declaration-of-helsinki/

When clinical trial participation and disease burden are not aligned geographically, benefits are distributed unequally. However, current frameworks do not systematically account for such misalignment. Existing indicators document disease burden concentration (where illness occurs) [24,25], funding allocation (where resources flow) [10], and publication patterns (where research is performed and disseminated) [26,27]. Yet no comprehensive framework measures whether populations bearing disease burden participate proportionally in research addressing that disease, both as research producers (where studies are conducted) and as enrolled participants (who access clinical trial benefits). This reflects a limitation in how we understand research inequality and points to the increasing demand of health policies to reduce epistemic injustice (i.e. knowledge practices that privilege dominant groups) in global health [28].

This has two major consequences. First, the lack of inclusion of low-income and of various population groups (such as women, children, elderly and those with common medical conditions) limits the generalizability of data guiding therapeutic interventions [29,30]. This is particularly concerning since excluded populations face disproportionately higher morbidity and mortality [31]. Second, RCTs are often based on convenience samples, affecting the extent to which they are representative of full populations [32,33]. Pharmaceutical companies value timely enrollment and data quality when allocating clinical trials, with study population availability and site resources being the most important [34]. These factors tend to lead RCTs toward established infrastructure rather than on populations with the highest burden, creating a self-reinforcing concentration of populations studied.

The paper is structured as follows. We first analyze 78,117 RCTs across 16 major disease categories conducted between 2000 and 2024, encompassing 185.6 million participants, and linked to time-varying Global Burden of Disease estimates for 180 countries. For each country-disease pair, we calculate the Participation-to-Burden Ratio (PBR) revealing systematic mismatches between where disease prevalence and research participants are from. Second, through variance decomposition, Shapley value analysis, and counterfactual removal techniques, we estimate whether participation inequality is primarily disease-driven or explained by country-level factors. Our analyses reveal that country-level factors account for most of the variance in participation patterns, whereas disease-specific factors contribute marginally. Third, we translate these findings into an actionable policy framework, which diagnoses for each country-disease pair whether research capacity, health infrastructure, or governance represents the primary limiting factor, enabling targeted interventions rather than generic capacity building recommendations. On the whole, our analysis suggests that addressing global health inequalities requires greater horizontal investments in research capacity, health infrastructure, and governance that operate across disease domains rather than within them.

# Results

## Geographic and income inequalities characterize participant enrollment

Clinical trial participation exhibits profound geographic and income-based inequalities that operate independently of disease burden, which aligns with WHO global observatory [34]. Countries in North America and Western Europe are consistently over-represented in RCTs enrollment across disease categories (**Extended Data Fig. 1 and 2**). For example, in cardiovascular diseases regions with lower disease burden exhibit the highest participation rates (**Fig. 1A**), with 71.9% of participants enrolled in the Global North, regardless of where disease is concentrated. Even for maternal and neonatal disorders, where the Global South bears 98.4% of the global disease burden, Global North countries represent 59% of RCTs participants (**Fig. 1B, Supplementary Table 6)**.

These geographic disparities manifest through country-level research specialization patterns that mirror income levels and regional divides rather than epidemiological needs. This is apparent when countries are grouped by research focus (i.e. specialization index, SI): North American and European countries form distinct groups separate from sub-Saharan African and South Asian countries (**Fig. 1D**). Most countries show below-average participation rates ($\text{Log(SI)} < 0$) across the majority of diseases, while several Global North countries (e.g. USA, Germany) maintain above-average rates across multiple disease categories simultaneously (**Fig. 1E**).

This pattern is even clearer when countries are grouped by income-level, as those tend to specialize in different disease categories (**Fig. 1C**). Several high-income countries show strong specialization in non-communicable diseases (i.e., high SI values), especially cardiovascular diseases, neurological disorders, and mental disorders. The clearest inequality by income level appears in the relationship between disease burden and RCTs participation (**Fig. 1F-I**). In high-income countries, disease burden is strongly associated with research participation ($\beta = 0.828$, $p < 0.001$), suggesting closer alignment between research activity and disease burden. This association weakens in middle-income countries ($\beta = 0.628\text{-}0.798$, $p<0.001$) and is weakest in low-income countries ($\beta = 0.278$, $p = 0.001$). The association is more than five times larger in high-income countries than in low-income countries (**Extended Data Fig. 2**). This indicates that RCTs participation aligns much more closely with disease burden in wealthier countries than in low-income countries.

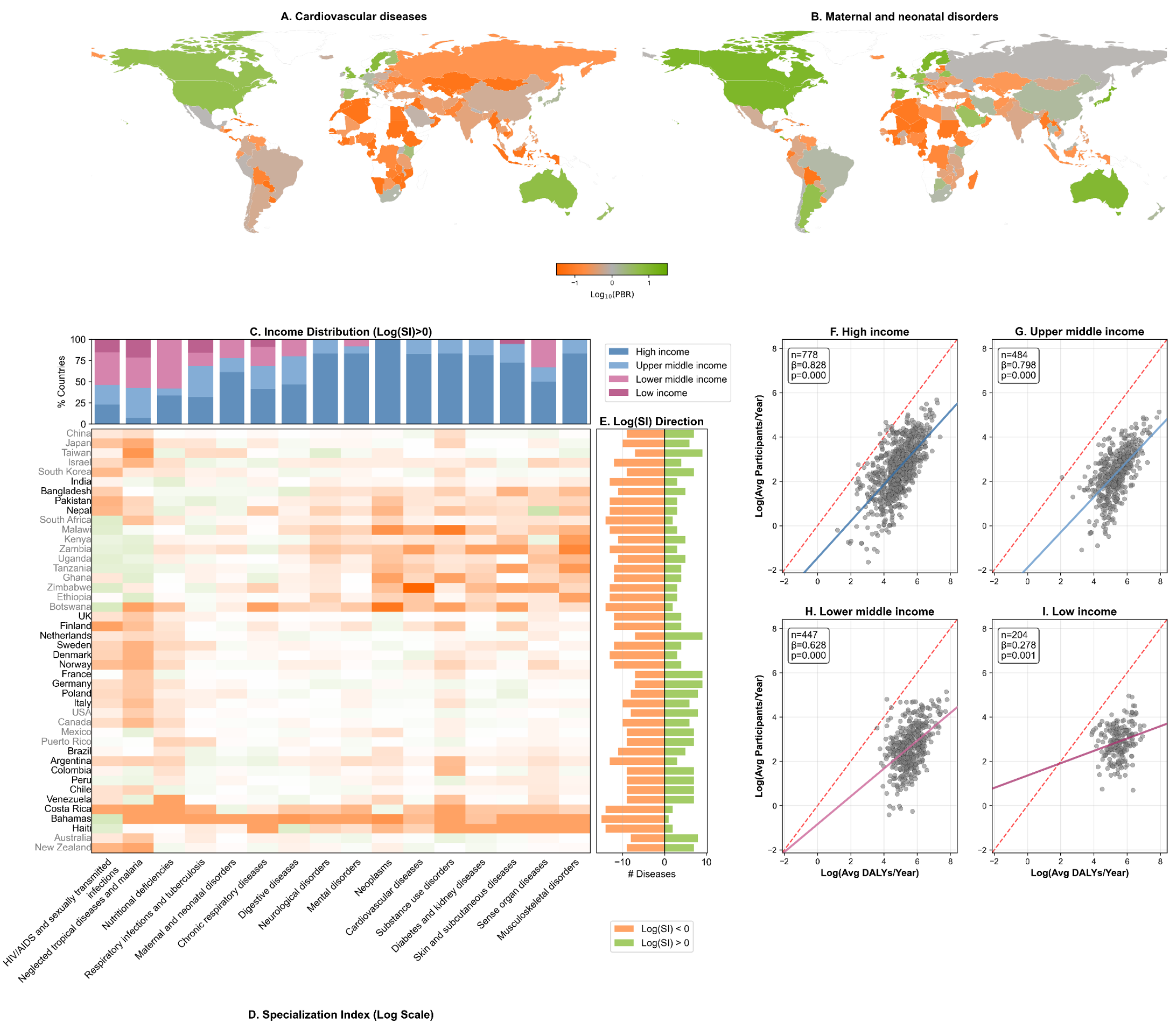

**Fig. 1. Geographic inequality in RCTs participation and specialization patterns.** Multi-panel figure examining geographic disparities in RCTs participation across diseases and income levels. **A-B:** world maps showing log-transformed participation-to-burden ratios (PBR) for cardiovascular diseases and HIV/AIDS respectively, using orange-gray-green colormap where orange indicates under-representation (LogPBR < 0), gray indicates proportional representation (LogPBR ≈ 0), and green indicates over-representation (LogPBR > 0). LogPBR scales from -1.5 to 1.5. World maps for other diseases are in **Extended Data Fig. 3**. **C:** stacked bar chart displaying income distribution of countries with positive specialization index (SI > 0) for each disease category, with bars colored by World Bank income classification. **D:** heatmap of log-transformed specialization index for top countries by region across Level 2 disease categories. **E:** stacked bar chart showing the count of diseases with positive versus negative SI for each country. **F-I:** scatter plots of log-transformed average annual DALYs (the total number of years of life lost because of illness, disability or premature death [10]) versus average annual participants for each disease-country pair, stratified by income level (high-income, upper middle income, lower middle income, low income). Gray dots represent individual disease-country pairs and colored lines show the fitted trend within each income group, and diagonal reference lines. Statistical annotations include sample size (n), regression coefficient (β), and p-values.

## Country-level factors account for most participation inequality

The geographic patterns observed across disease categories suggest that differences in research participation are larger across countries than those observed across diseases. To quantify the relative contribution of each of these two dimensions, we applied a series of decomposition and sensitivity analyses. We evaluated, for each disease category, its marginal contribution to global inequality by quantifying the change in the overall Gini coefficient when specific disease's Gini coefficient was excluded. Across all 16 Level 2 disease categories, contributions to inequality score (CIS, calculation in **Supplementary Methods 3**) are consistently small (**Fig. 2A**). Even the largest contributors, neoplasms and maternal and neonatal disorders, account for only 1.0% and 0.6% of total inequality, respectively. Several diseases traditionally considered underfunded, including nutritional deficiencies, exhibited negative CIS values, indicating that their research-burden alignment slightly lowers the overall inequality score. These results demonstrate that no single disease, nor any small subset of diseases, drives global participation inequality.

In contrast, geographic structure shows markedly stronger effects. Variance partitioning of PBR reveals that country factors account for almost all of the explanatory power. In a two-way decomposition analysis (**Supplementary Methods 3)**, country accounts for 88.9% of the total variance (partial $R^2$ = 0.439), whereas disease explains only 3.4% (partial $R^2$ = 0.017), and temporal effects accounts for the remaining 7.7% (partial $R^2$ = 0.038) (**Extended Data Fig. 4B**). Country effects are thus almost 30 times larger than disease effects, indicating that participation inequality is associated much more with country differences than with disease focus.

Decomposition of temporal inequality trends further supports this finding. Differences in average RCTs participation across diseases declined from 40% to 29% in 25 years (**Fig. 2C**), suggesting convergence across diseases over time. In contrast, inequality within diseases, which captures geographic disparities in participation for a given disease, increased from 59% to 88% in 2018 and remained high in subsequent years, indicating that participation inequality became more closely associated with country differences than with disease over time.

Additional sensitivity analyses based on removal confirmed this trend. Excluding the top 20% of diseases ranked by CIS (3 of 16) reduces the global Gini coefficient by only 1.9%. By contrast, removing the top 20% of countries ranked by RCTs participation volume (36 of 180) reduces inequality by 19.1%, an effect that is more than 10 times larger. Consistent with this asymmetry, the country-level Lorenz curve with higher Gini exhibits a pronounced shift after the exclusion of major contributing countries (Gini: 0.879 to 0.711), whereas removing diseases produces only a negligible change (Gini: 0.801 to 0.786) (**Fig. 2B and 2D**).

Together, these findings clearly show that country-level factors account for most of the observed inequality in research participation. Targeted interventions organized around specific diseases may influence participation for particular conditions, such as HIV/AIDS research, but such effects operate within a wider geographic context that shapes participation across all disease domains. Regardless of disease burden or epidemiological profile, research participation follows a consistent global hierarchy influenced primarily by country-level factors.

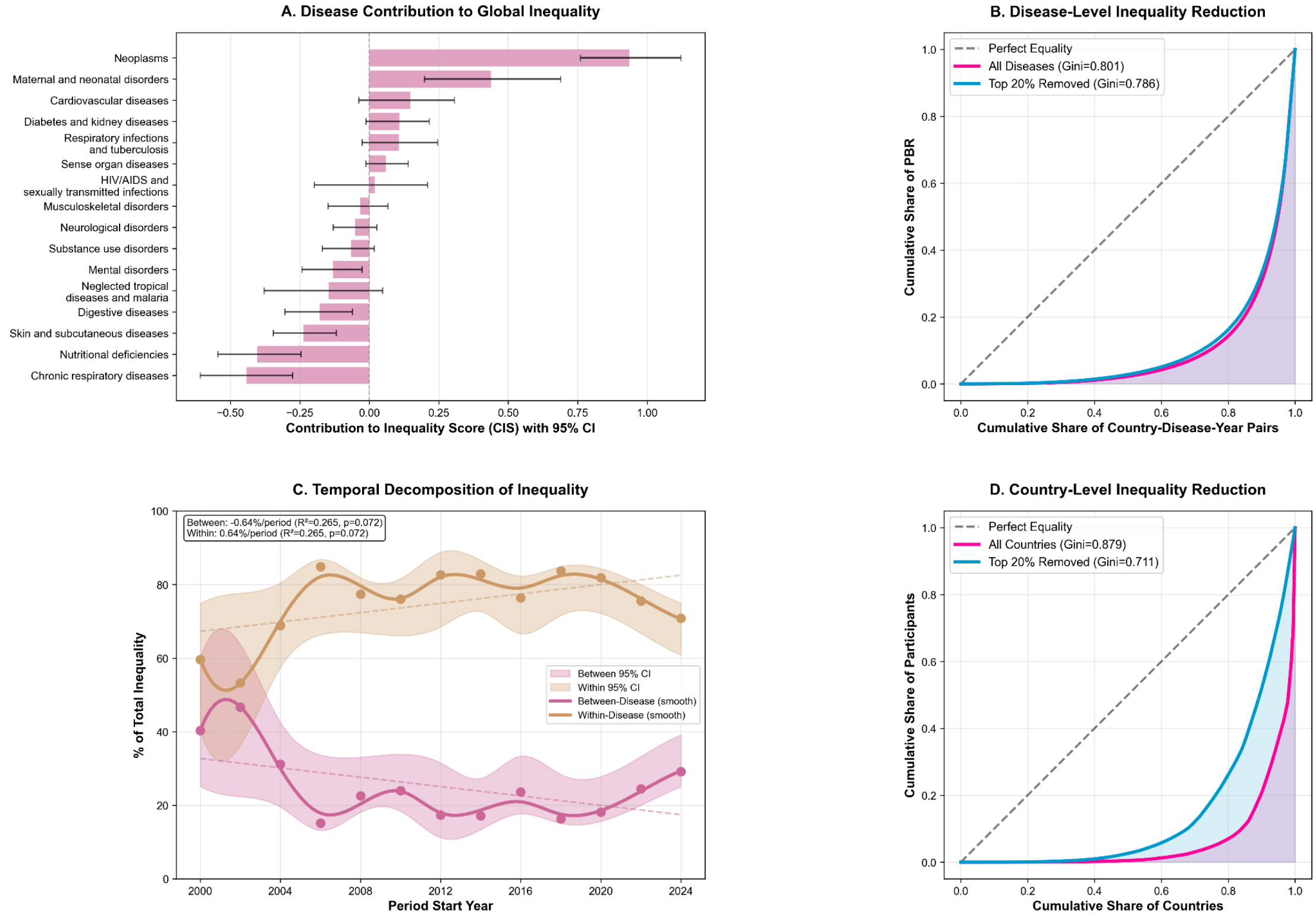

**Fig 2. Decomposition analysis of participation inequality (2000-2024). A:** Contribution to Inequality Score (CIS) for each Level 2 disease category, with 95% bootstrap confidence intervals, showing small contributions from individual disease categories to global inequality. **B:** Lorenz curves comparing observed inequality with all diseases included versus after excluding the top 20% of diseases by CIS, illustrating minimal sensitivity to disease exclusion. **C:** Temporal decomposition of inequality into between-disease and within-disease components, showing declining between-disease inequality and increasing within-disease inequality over time. **D:** Country-level Lorenz curves comparing observed inequality with all countries included versus after excluding the top 20% of countries by participation volume, showing greater sensitivity to country removal.

## Unveiling specific country-level factors

To better understand why RCTs participation is concentrated in certain countries, we examine economic conditions, research capacity, health infrastructure, and governance and societal factors (**Supplementary Table 7 and Extended Data Fig. 5**). The findings show that wealth and population size explain a large share of country-level variation in participation-to-burden ratios. The 15-variable model accounts for 50.3% of the total variation in logPBR (**Supplementary Table 8**). Using variable-level Shapley value decomposition to partition this explained variance, we find that economic factors (population size, GDP, and foreign aid) account for the largest share (51.7%), followed by research capacity (R&D expenditure, total publications, citations, and researchers per million) for 30.8%. Population size alone contributes the largest share (30.4%). This finding is intuitive: countries with large populations and sizable economies have greater capacity to conduct RCTs, as they provide more stable research ecosystems and better access to large patient pools. These factors represent the basic conditions for conducting RCTs, but they change slowly, they create persistent asymmetries that are difficult to overcome through short-term or targeted interventions.

While economic factors account for the largest share of country-level inequality, isolating residual variation helps identify areas where health policy can act. Research capacity, health infrastructure (health expenditure per capita, UHC index, medical schools, and sanitation access), and governance (HDI, democracy index, altruism score, and trust in scientists) together explain 24.8% of the remaining unexplained residual variance (**Supplementary Table 9**). Although smaller in magnitude than economic factors, their influence is meaningful as they point to specific areas for policy actions to strengthen countries' research and health systems. Among those, research capacity is the most influential in reducing inequality in RCTs participation. Research capacity accounts for 15.4% of the remaining residual variance, while health infrastructure adds another 8.3%. This underscores how both R&D workforce, publication output, scientific impact and clinical facilities (such as medical schools and sanitation access) influence the feasibility of conducting RCTs once economic factors are setting aside. Governance and societal factors contribute a small portion (1.2%), highlighting that institutional stability and social trust play a supporting role in optimizing how capacity translates into active participation.

These findings show that research inequalities are rooted in economic conditions, which give wealthy countries clear advantages in conducting RCTs. Nevertheless, there is room for targeted policy interventions, as research capacity, health infrastructure, and governance explain meaningful differences in participation even after excluding economic factors.

**Horizontal interventions to reduce participation inequality**

Our analysis provides strong evidence that global inequality in RCTs participation is associated far more with country-level factors than with disease burden. Even setting aside economic conditions, participation inequality remains linked to research capacity, health infrastructure, and governance. These findings provide a concrete basis for reducing participation inequality at the country-level, consistent with the core principle of horizontal intervention: strengthening the research and health systems conditions that operate across diseases through coordination and collaboration between funders, health actors and policymakers [36].

To explore opportunities for horizontal interventions, we identified research, health, and governance factors associated with inequality for 1,962 country and disease pairs using linear regression. We then classified these pairs into four performance types based on deviations from expected participation levels: Under-performing (557 pairs), Over-performing (331 pairs), As-expected (170 pairs), and Intermediate (904 pairs). These types reflect different patterns of global inequality: Under-performing pairs show little global inequality contribution, indicating that they are primarily affected by low participation relative to disease burden rather than contributing to global inequality (**Fig. 3A**). Over-performing pairs exhibit both large positive residuals and high CIS, indicating they contribute significantly to inequality (e.g., diabetes and kidney diseases-FIN: CIS=1.461) (**Fig. 3B**). Extreme PBR values are concentrated in five countries: Finland, Norway, Sweden, UK, and Denmark have PBRs higher than 10, whereas the remaining 175 countries range from 0.000002 to 8.9 (**Supplementary Data**). Comparing factor associations across the four performance types **(Fig. 3A-C)** help explain these differences. Research capacity is key: under-performing pairs struggle with increasing research capacity, while over-performing pairs benefit from greater access to research capacity and governance. However, neither of them shows a strong influence by health infrastructure. Interestingly, as-expected pairs characterized by stronger health infrastructure are associated with more balanced participation rate than the other two patterns.

We further examined how factors and performance types are connected globally by visualizing the country-factor-disease network (**Fig. 3D**). First, research capacity is the most influential factor across all performance types, being connected to the broadest range of diseases (3.20 diseases per node; **Supplementary Tables 12B**) and connecting 37.5% of under-performing pairs. Second, governance and societal factors are linked to fewer diseases (2.29 on average) and appear relatively isolated from other factors. Third, health infrastructure occupies a brokerage position, connecting to a relatively broad number of diseases (2.64 on average) and bridging research capacity and governance and societal factors. Notably, only 8.8% of country-disease pairs are affected by more than one factor, indicating that most participation misalignments are linked to one specific research, health or governance factor.

The lack of research capacity among under-performing country-disease pairs, together with the disproportionate contribution of over-performing pairs to inequality, points to where action is needed. However, the question of which type of action is more effective to reduce inequality remains open. We tested two courses of action using simulated scenarios: (1) Full Alignment (all countries shift toward median PBR), representing maximum theoretical equality; and (2) Targeted Alignment (only the most misaligned countries adjusted), reflecting a more realistic approach. Full alignment would eliminate all avoidable inequality (Gini reduction: 100% from 0.772 to 0.000). Targeted alignment would be remarkably more efficient: adjusting just the top 40% of countries (72 of 180) would reduce inequality by 58.4%, while adjusting only the top 10% (18 countries) would achieve 22.2% reduction (**Fig. 3E, Supplementary Tables 13**). Targeted alignment is 1.46 times more efficient per country adjusted than full alignment (**Supplementary Tables 13**), showing that extreme misalignment is concentrated in a subset of countries where substantial reduction could be achieved by focusing efforts on the most misaligned countries.

Both scenarios point to horizontal interventions among countries facing comparable constraints, enabling coordinated action, shared infrastructure, and joint research efforts across disease areas. We modeled the effects of the two simulated scenarios on the global research collaboration network (450 nodes, 25,047 edges). Under full alignment, network density increases by 50% (from 0.248 to 0.372), while homophily, defined as connections among nodes sharing the same factor, decreases by 39.8% (from 0.457 to 0.275), indicating stronger network cohesion but also greater opportunities for collaboration and coordinated action as countries share similar problems (**Fig. 3F**). Targeted alignment produces a similar pattern of integration: density increases by 0.364 (46.8%), while homophily decreases by 0.287 (37.2%) (**Supplementary Tables 14**). These shifts show that horizontal interventions based on coordinated efforts among countries can both reduce participation inequality and foster a more cohesive and diverse global research ecosystem.

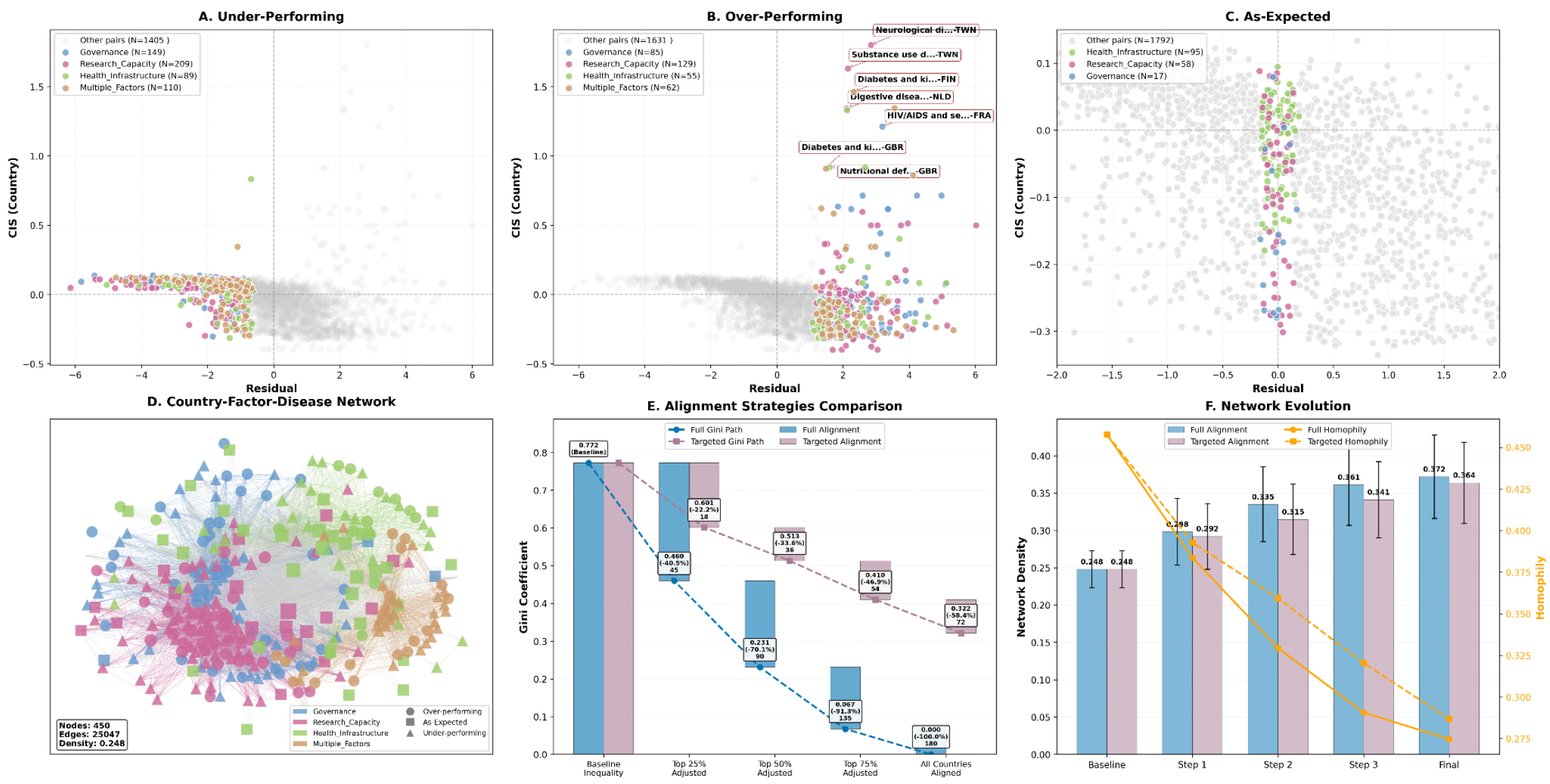

**Fig. 3. Simulated interventions to reduce inequality across country-disease pairs**. **A:** Scatter plot of Under-performing showing residual (x-axis) versus CIS (y-axis). Points are colored by factor dominance (Research Capacity, Health Infrastructure, Governance, Multiple Factors, Other Pairs). **B:** Scatter plot of Over-performing (same display as A). C**:** Scatter plot of As-expected (same display as A). **D:** Network visualization showing connections between country-factor pairs (nodes). Nodes are colored by dominant factor, node shapes indicate performance type, node size is proportional to disease count, and edge thickness represents the weight of shared disease connections. **E:** Waterfall chart showing the reduction in the Gini coefficient under two scenarios of interventions. X-axis: Baseline, top 25%-50%-75%-100%, adjusted for Full Alignment, and Baseline, Top 10%-20%-30%-40%, for Targeted Alignment. The y-axis shows the Gini coefficient. Bars represent the Gini reduction at each step. Text labels show the Gini value, percentage reduction, and number of countries adjusted. **F:** Network metrics under full and targeted alignment. Bars show Network Density for Full Alignment (blue) and Targeted Alignment (pink) scenarios with error bars. Lines show Homophily (proportion of same-factor connections) reduction for both scenarios. The x-axis represents intervention steps matching panel E. Right Y-axis shows Homophily values. Summary text indicates percentage changes in density and cross-factor connections. Note that in panel E, the calculation of Gini coefficient differs from Fig. 2B and 2D because different calculation subjects are used.

## Discussion

Our analyses demonstrate that clinical trial participation exhibits persistent and pronounced inequalities that are not explained by the burden of disease, but rather reflects a structural property of the global health system. Across disease areas, participation patterns are explained primarily by country-level factors and remain stable over time, despite intensified efforts to address disease burden [37]. This decoupling indicates that vertical programs, that is efforts organized around specific diseases and delivered through targeted interventions, have had little effect in mitigating inequalities in clinical trial participation. Our simulations show that changing this unequal global structure of participation requires greater horizontal investments to strengthen health research systems, research capacity and governance across disease areas, confirming the initiatives from previous works [4, 48].

The dominance of country-level effects over disease-specific factors also indicates that clinical trial participation is primarily influenced by the location where clinical trials can be conducted rather than where the burden of disease is greatest. Clinical trial populations represent the empirical basis for evaluating medical safety, efficacy, and comparative effectiveness [33], and prior work has shown that clinical outcomes, treatment responses, and comorbidity profiles vary across populations and health systems [25,30,38-41]. Our findings quantified this by showing the clinical evidence base is systematically biased toward participants from Global North. Financial resources to conduct clinical trials are often mobilized through national health systems and domestic constraints (regulatory regimes, institutional administration, infrastructure), which limits opportunities for countries with lower capacity [17]. Our findings indicate that research capacity exerts a more substantial influence on participation than health infrastructure, underscoring its central role in addressing global inequities. While disease-specific initiatives continue to be critical for establishing care infrastructure in high-burden areas, as seen in HIV/AIDS in sub-Saharan Africa, they serve a distinct role from systemic capacity building. Nevertheless, the participation misalignment cannot be addressed at the level of countries alone as modern clinical trials operate within transnational networks of sponsors, sites, and complex regulations [38,42]. Thus, intervention requires coordinated efforts to change how the global research system allocates opportunities to participate in, and thereby shape, the production of medical evidence [43]. Models of coordinated funding have recently emerged that can help direct resources toward countries with lower health and research capacity [44].

Addressing participation inequality is crucial for scientific and ethical reasons. A central feature of clinical trials is that their results can be reasonably applied to a well-defined population [44]. If the evidence produced is disproportionately built from participants in lower-burden countries or populations with limited diversity, the evidence on the effects of the intervention may be limited precisely for the groups most likely to benefit from the clinical trial outcome. This concern is well-documented in studies examining the implementation of clinical trial findings in external settings, which show that treatment effectiveness can vary with health system capacity, background risk, and population characteristics [45-47]. Furthermore, reshaping participation is not only a distributional issue but also a matter of epistemic justice, as it implies recognizing populations usually marginalized in global health research as valid producers and recipients of knowledge [28,43].

Targeted funding can increase research on diseases with a high burden in low-income countries. However, our results indicate that participation remains concentrated in high-income countries, where conditions are more favorable for conducting clinical trials and enrolling participants [48,49]. Thus, changing the unequal structure of clinical trial participation would require greater emphasis in horizontal interventions. This interpretation is consistent with studies showing that external funding translates into outcomes only where absorptive capacity is sufficient, highlighting the need for horizontal investments that build that capacity [50,51]. Vertical programs based on disease-targeted interventions have dominated health funding, despite previous calls to reconcile both approaches [17]. This dominance can be explained by considerations of cost-effectiveness in impact evaluations, the status of randomized control trials as the golden standard [21], and the search for single interventions that are feasible for donors and governments. Building infrastructure or developing integrated systems that operate across disease areas can be costly, less suited to randomization, and more difficult to assess in terms of impact (e.g. lives saved) [17,36]. Despite these challenges, increasing support for horizontal interventions is crucial for improving equity in global health [31,52] as funding allocation strategies based on disease-targeted interventions are unable to substantially alter where clinical research activity occurs.

Our research has limitations. First, it is based on published RCTs that are indexed in PubMed, which allows us to analyze enrollment locations, sample sizes, and disease focus, with an emphasis on relevance and accuracy. Such sampling, however, may underrepresent null-result because of publication bias, as well as early-phase studies, even though it has been complemented with data from ClinicalTrials.gov (pre-registrations are less consistent in reporting participant geography and are often not linked to published articles [53]). Second, participation measures at the macro level cannot capture all dimensions of clinical trial involvement, especially with missing data from traditionally underestimated low-income countries. These limitations, however, do not affect our findings, which remain stable across alternative classification thresholds, disease exclusions, and decomposition strategies, with country effects consistently accounting for the majority of the observed variation. Third, while we include a multidimensional set of national and institutional factors, these indices still capture only selected dimensions of capacity and may not fully reflect the complexities of national and institutional arrangements. Future research could further refine these measures to better account for the nuanced interplay between formal policy and informal social norms.

Reducing participation inequality in clinical trials demands reconsidering prevailing approaches to global health. Taken together, our findings suggest three priority areas for action. First, changing entrenched patterns of participation inequality requires greater horizontal investment in developing research capacity to conduct clinical trials, but also in building and developing health systems and infrastructure that enable them. This will require both an increase in the research capacity of low- and middle-income countries, and sustained support from international funders, particularly where national resources remain limited. A diagonal approach [19] can align vertical and horizontal interventions so that funding directs toward diseases with a high burden and also strengthens research capacities that can be used more broadly. Second, effective interventions should be tailored to similar country-disease contexts rather than applying uniform capacity-building solutions. Third, coordinated and collaborative efforts that redistribute research capacity and epistemic authority are essential to ensure that participation yields shared knowledge production, while future efforts broader health system strengthening and long-term therapeutic availability. Addressing global health inequalities, therefore, requires treating research participation as a structural property of the global health system, not simply as a disease-specific problem.

## References

[1] Apeagyei, A. E., Bisignano, C., Elliott, H., Hay, S. I., Lidral-Porter, B., Nam, S., ... & Dieleman, J. L. (2025). Tracking development assistance for health, 1990–2030: historical trends, recent cuts, and outlook. *The Lancet*, 406(10501), 337-348.

[2] Micah, A. E., Bhangdia, K., Cogswell, I. E., Lasher, D., Lidral-Porter, B., Maddison, E. R., ... & Hlongwa, M. M. (2023). Global investments in pandemic preparedness and COVID-19: development assistance and domestic spending on health between 1990 and 2026. *The Lancet Global Health*, 11(3), e385-e413.

[3] Dieleman, J., Campbell, M., Chapin, A., Eldrenkamp, E., Fan, V. Y., Haakenstad, A., ... & Murray, C. J. (2017). Evolution and patterns of global health financing 1995–2014: development assistance for health, and government, prepaid private,

and out-of-pocket health spending in 184 countries. *The Lancet*, 389(10083), 1981-2004.

[4] Ratevosian, J., Millett, G., Honermann, B., Bennett, S., Connor, C., Bekker, L. G., & Beyrer, C. (2025). PEPFAR under review: what's at stake for PEPFAR's future. *The Lancet*, 405(10479), 603-605.

[5] Rad, J. (2025). Health inequities: a persistent global challenge from past to future. *International Journal for Equity in Health*, 24(1), 148.

[6] De Maio, F. (2014). *Global health inequities: A sociological perspective*. Bloomsbury Publishing.

[7] Elendu, C., Amaechi, D. C., Elendu, T. C., Amaechi, E. C., Elendu, I. D., Akpa, K. N., ... & Idowu, O. F. (2025). Shaping sustainable paths for HIV/AIDS funding: a review and reminder. *Annals of Medicine and Surgery*, 87(3), 1415-1445.

[8] Baru, R. V., & Mohan, M. (2018). Globalisation and neoliberalism as structural drivers of health inequities. *Health Research Policy and Systems*, 16(Suppl 1), 91.

[9] Cash-Gibson, L., Rojas-Gualdrón, D. F., Pericàs, J. M., & Benach, J. (2018). Inequalities in global health inequalities research: A 50-year bibliometric analysis (1966-2015). *PloS one*, 13(1), e0191901.

[10] Schmallenbach, L., Bley, M., Bärnighausen, T. W., Sugimoto, C. R., Lerchenmüller, C., & Lerchenmueller, M. J. (2025). Global distribution of research efforts, disease burden, and impact of US public funding withdrawal. *Nature Medicine*, 1-9.

[11] Myers, K. (2020). The Elasticity of Science. *American Economic Journal: Applied Economics*, 12(4), 103–134.

[12] Barrenho, E., Miraldo, M., & Smith, P. C. (2019). Does global drug innovation correspond to burden of disease? The neglected diseases in developed and developing countries. *Health Economics*, 28(1), 123–143.

[13] Yegros-Yegros, A., van de Klippe, W., Abad-Garcia, M. F., & Rafols, I. (2020). Exploring why global health needs are unmet by research efforts: the potential influences of geography, industry and publication incentives. *Health Research Policy and Systems*, 18(1), 47.

[14] Charpignon, M. L., Matos, J., Nakayama, L. F., Gallifant, J., Alfonso, P. G. I., Cobanaj, M., ... & Celi, L. A. (2025). Diversity in the medical research ecosystem: a descriptive scientometric analysis of over 49 000 studies and 150 000 authors published in high-impact medical journals between 2007 and 2022. *BMJ open*, 15(1), e086982.

[15] Payedimarri, A.B., Mouhssine, S., Aljadeeah, S. et al. (2026). Assessing patterns of authorship of low- and middle-income countries in global commercial clinical trials in oncology. Global Health 22, 3.

[16] Chang, A. Y., Cowling, K., Micah, A. E., Chapin, A., Chen, C. S., Ikilezi, G., ... & Qorbani, M. (2019). Past, present, and future of global health financing: a review of development assistance, government, out-of-pocket, and other private spending on health for 195 countries, 1995–2050. *The Lancet*, 393(10187), 2233-2260.

[17] Béhague, D. P., & Storeng, K. T. (2008). Collapsing the vertical–horizontal divide: An ethnographic study of evidence-based policymaking in maternal health. *American Journal of Public Health*, 98(4), 644-649.

[18] De Maeseneer, J., Van Weel, C., Egilman, D., Mfenyana, K., Kaufman, A., & Sewankambo, N. (2008). Strengthening primary care: addressing the disparity between vertical and horizontal investment. *The British Journal of General Practice*, 58(546), 3.

[19] Kirwin, E., Meacock, R., Round, J., & Sutton, M. (2022). The diagonal approach:

A theoretic framework for the economic evaluation of vertical and horizontal interventions in healthcare. *Social Science & Medicine*, 301, 114900.
[20] Zhao, G., Cao, X., & Ma, C. (2020). Accounting for horizontal inequity in the delivery of health care: A framework for measurement and decomposition. *International Review of Economics & Finance*, 66, 13-24.
[21] Bothwell, L. E., Greene, J. A., Podolsky, S. H., & Jones, D. S. (2016). Assessing the gold standard-lessons from the history of RCTs. *New England Journal of Medicine*, 374(22), 2175-2181.
[22] Okpechi, I. G., Swanepoel, C. R., & Venter, F. (2015). Access to medications and conducting clinical trials in LMICs. *Nature Reviews Nephrology*, 11(3), 189-194.
[23] Hay, S. I., Ong, K. L., Santomauro, D. F., Aalipour, M. A., Aalruz, H., Ababneh, H. S., ... & Ajose, A. O. (2025). Burden of 375 diseases and injuries, risk-attributable burden of 88 risk factors, and healthy life expectancy in 204 countries and territories, including 660 subnational locations, 1990–2023: a systematic analysis for the Global Burden of Disease Study 2023. *The Lancet*, 406(10513), 1873-1922.
[24] Lohr, K. N., Eleazer, K., & Mauskopf, J. (1998). Health policy issues and applications for evidence-based medicine and clinical practice guidelines. *Health Policy*, 46(1), 1-19.
[25] Singh, L., Cristia, A., Karasik, L. B., Rajendra, S. J., & Oakes, L. M. (2023). Diversity and representation in infant research: Barriers and bridges toward a globalized science of infant development. *Infancy*, 28(4), 708-737.
[26] Evans, J. A., Shim, J. M., & Ioannidis, J. P. (2014). Attention to local health burden and the global disparity of health research. *PloS one*, 9(4), e90147.
[27] Zhou, H., Garg, P., & Fetzer, T. (2025). The Changing Geography of Medical Research. *medRxiv*, 2025-09.
[28] Bhakuni, H., & Abimbola, S. (2021). Epistemic injustice in academic global health. *The Lancet Global Health*, 9(10), e1465–e1470.
[29] Van Spall, H. G., Toren, A., Kiss, A., & Fowler, R. A. (2007). Eligibility criteria of randomized controlled trials published in high-impact general medical journals: a systematic sampling review. *JAMA*, 297(11), 1233-1240.
[30] Sugimoto, C. R., Ahn, Y. Y., Smith, E., Macaluso, B., & Larivière, V. (2019). Factors affecting sex-related reporting in medical research: a cross-disciplinary bibliometric analysis. *The Lancet*, 393(10171), 550-559.
[31] Warner, E., Marron, J. M., Peppercorn, J. M., Abel, G. A., & Hantel, A. (2024). Shifting from equality toward equity: addressing disparities in research participation for clinical cancer research. *The Journal of Clinical Ethics*, 35(1), 8-22.
[32] Clougherty, J. E., Kinnee, E. J., Cardet, J. C., Mauger, D., Bacharier, L., Beigelman, A., ... & Holguin, F. (2021). Geography, generalisability, and susceptibility in clinical trials. *The Lancet Respiratory Medicine*, 9(4), 330-332.
[33] Rothwell, P. M. (2005). External validity of randomised controlled trials: “to whom do the results of this trial apply?”. *The Lancet*, 365(9453), 82-93.
[34] World Health Organization: WHO. (2023, November 21). Global Observatory on Health R&D: Bridging the Gap in Global Health Research and Development. /. https://www.who.int/news/item/21-11-2023-global-observatory-on-health-r-d--bridging-the-gap-in-global-health-research-and-development.
[35] Goldacre, B., Lane, S., Mahtani, K. R., Heneghan, C., Onakpoya, I., Bushfield, I., & Smeeth, L. (2017). Pharmaceutical companies’ policies on access to trial data, results, and methods: audit study. *BMJ*, 358.

[36] Stoljar Gold, A. (2025). Vertical dominance: Cost-effectiveness, randomisation, and the bias against horizontal interventions in global health. *Global Public Health*, 20(1), 2523542.
[37] World Health Organization. (2025). *A global health strategy for 2025-2028- advancing equity and resilience in a turbulent world: fourteenth General Programme of Work*. World Health Organization.
[38] Djurisic, S., Rath, A., Gaber, S., Garattini, S., Bertele, V., Ngwabyt, S. N., ... & Gluud, C. (2017). Barriers to the conduct of randomised clinical trials within all disease areas. *Trials*, 18(1), 360.
[39] Yaffe, K., Vittinghoff, E., Dublin, S., Peltz, C. B., Fleckenstein, L. E., Rosenberg, D. E., ... & Larson, E. B. (2024). Effect of personalized risk-reduction strategies on cognition and dementia risk profile among older adults: the SMARRT randomized clinical trial. *JAMA* Internal Medicine, 184(1), 54-62.
[40] Eckey, M., Li, P., Morrison, B., Bergquist, J., Davis, R. W., & Xiao, W. (2025). Patient-reported treatment outcomes in ME/CFS and long COVID. *Proceedings of the National Academy of Sciences*, 122(28), e2426874122.
[41] Yoo, S. K., Fitzgerald, C. W., Cho, B. A., Fitzgerald, B. G., Han, C., Koh, E. S., ... & Chowell, D. (2025). Prediction of checkpoint inhibitor immunotherapy efficacy for cancer using routine blood tests and clinical data. *Nature Medicine*, 31(3), 869-880.
[42] Zerhouni, E. A. (2005). Translational and Clinical Science--Time for a New Vision. *New England Journal of Medicine*, 353(15), 1621-1623.
[43] Sumathipala, A., Jayasinghe, O. S., & Fernando, B. (2025). Key to successful global health collaborations: research, ethics and community engagement and involvement. *BMJ leader*, 9(2).
[44] World Health Organization. (2024). *Guidance for best practices for clinical trials. In Guidance for best practices for clinical trials*. https://iris.who.int/server/api/core/bitstreams/dbe6d97c-b659-4e7b-8070-b4a3c667d6e5/content.
[45] Vaz, L. M., Franco, L., Guenther, T., Simmons, K., Herrera, S., & Wall, S. N. (2020). Operationalising health systems thinking: a pathway to high effective coverage. *Health Research Policy and Systems*, 18(1), 132.
[46] Nordon, C., Karcher, H., Groenwold, R. H., Ankarfeldt, M. Z., Pichler, F., Chevrou-Severac, H., ... & Abenhaim, L. (2016). The "efficacy-effectiveness gap": historical background and current conceptualization. *Value in Health*, 19(1), 75-81.
[47] Dorresteijn, J. A., Visseren, F. L., Ridker, P. M., Wassink, A. M., Paynter, N. P., Steyerberg, E. W., ... & Cook, N. R. (2011). Estimating treatment effects for individual patients based on the results of randomised clinical trials. *BMJ*, 343.
[48] Kennedy, A., & Ijsselmuiden, C. (2008). Country ownership and vertical programmes in health, health information and health research. B*ulletin of the World Health Organization*, 86(8), C–D.
[49] COHRED Statement 1: Responsible vertical programming of global health. (2012, June 22). Council on Health Research for Development - COHRED. https://www.cohred.org/publications/cohred-publications/policy-and-synthesis/cohred-statement-1-responsible-programming-of-global-health-research/
[50] Stephan, P. E. (1996). The economics of science. *Journal of Economic Literature*, 34(3), 1199-1235.
[51] Yin, Y., Dong, Y., Wang, K., Wang, D., & Jones, B. F. (2022). Public use and

public funding of science. *Nature Human Behaviour*, 6(10), 1344-1350.

[52] Aiyegbusi, O. L., Cruz Rivera, S., Kamudoni, P., Anderson, N., Collis, P., Denniston, A. K., ... & Calvert, M. J. (2024). Recommendations to promote equity, diversity and inclusion in decentralized clinical trials. *Nature Medicine*, 30(11), 3075-3084.

[53] Chen, R., Desai, N. R., Ross, J. S., Zhang, W., Chau, K. H., Wayda, B., ... & Krumholz, H. M. (2016). Publication and reporting of clinical trial results: cross sectional analysis across academic medical centers. *BMJ*, 352.

## Methods

### Data sources, sample, and indicators

We identified randomized controlled trials (RCTs) through PubMed using validated search filters for human studies and clinical trial publication types. The initial retrieval yielded 301,262 documents published between 1980 and 2024. Following representativeness analysis, which revealed systematic under-representation of geographic information in pre-2000 publications (geographic annotation success rate: 38% for 1980-1999 versus 72% for 2000-2024), we restricted analysis to 2000-2024, yielding 193,806 eligible studies. Post-hoc power analysis confirmed adequate sample size to detect effect sizes ≥0.3 with 80% power at α=0.05（**Supplementary Methods 6**).

Disease burden data were obtained from the Global Burden of Disease data base 2023, maintained by the Institute for Health Metrics and Evaluation (IHME) at University of Washington [54], which provides disability-adjusted life years (DALYs) by country, year, and cause through hierarchical disease categorization. We focused on GBD Level 2 categories, which balance granularity with interpretability at the global macro level. Following previous research [13,26], causes that are difficult to assign to specific diseases (for example, "other non-communicable diseases") were excluded. We integrated disease burden associations from lower hierarchical levels (Levels 3 and 4) and aggregated them to Level 2, ensuring comprehensive coverage. In total, we examined 16 Level 2 disease categories, encompassing both communicable diseases (HIV/AIDS and sexually transmitted infections, neglected tropical diseases and malaria, respiratory infections and tuberculosis, maternal and neonatal disorders, nutritional deficiencies) and non-communicable diseases (cardiovascular diseases, neoplasms, diabetes and kidney diseases, chronic respiratory diseases, digestive diseases, mental disorders, neurological disorders, musculoskeletal disorders, sense organ diseases, skin and subcutaneous diseases, substance use disorders). The category Enteric infections was removed from analysis due to insufficient publication data.

Country-level indicators were compiled from multiple authoritative sources with temporal alignment to study periods. We included 15 factors that account for four critical categories of country-level inequality (**Supplementary Tables 7**): Economic (GDP per capita, population, foreign aid received), Research Capacity (R&D expenditure, publications, citations, researchers per million), Health Capacity (health expenditure per capita, UHC index, medical schools, sanitation), and Governance and societal context (HDI, democracy index, altruism, trust in scientists).

All datasets were harmonized across three dimensions-geography (country-level ISO3 codes), time (publication year), and medical domain (mapped disease categories) to enable analysis.

### Geographic annotation and participants extraction

Geography information was characterized across two dimensions: authorship location and participant enrollment sites. Author affiliations were extracted from PubMed metadata, which records institutional affiliations for corresponding and first authors. For multi-country studies, we applied a full counting, distributing credit proportionally across contributing countries.

Participant enrollment geography, which is not reported in a standardized way in publication databases, was extracted using a two-stage AI-assisted extraction pipeline. We initially screened 301,262 PubMed publications in the first round; we benchmarked four methods (string matching, spaCy, LLaMA, and Gemma) on 359 publications (**Supplementary Methods 1**). After finding optimal precision and recall (**Supplementary Table 3, Extended Data Fig. 6**), we implemented a second round using 1,482 stratified samples. We tested Gemini 3.5 Flash across three types of data input, and selected the cascade method (Title and Abstract + Participation + Methods + Results + Rest sections), later applied to 193,806 publications, resulting in 111,407 publications with enrollment location and amount (**Supplementary Table 4, Extended Data Fig. 7**). We complemented the above data with ClinicalTrials.gov data (36,687 publications are linked). After manual check and duplicate removal, 124,189 publications were ready to integrate with disease information (**Extended Data Fig. 5**).

Geographic data identified in publications were standardized to ISO3 country codes using validated mapping dictionaries (**Extended Data Fig. 5**). For studies reporting enrollment across multiple countries, participants were distributed proportionally based on their population across mentioned countries. Participant information was extracted using rule-based patterns validated through manual review of 500 articles, achieving 98.6% accuracy for reported sample sizes. Studies lacking geographic annotation or participant count information after extraction were excluded from analysis. Our final analytical dataset comprised 78,117 studies with complete geographic, participant, and disease information, representing 185.4 million RCTs participants.

**Disease classification and medical concepts harmonization**

Linking clinical trial publications to epidemiologically defined disease categories requires harmonizing heterogeneous medical classification systems that were developed for distinct purposes. International Classification of Diseases (ICD) codes are designed for clinical documentation and administrative reporting, whereas Global Burden of Disease (GBD) cause categories are designed to measure and compare disease burden across populations. Prior studies [55,56] have shown that direct mapping between these classification systems is limited, due to mismatches in scope and granularity.

To address this challenge, we adopted a conservative, ontology-mediated harmonization strategy that prioritizes semantic validity over maximal coverage **(Supplementary Methods 2)**. Instead of mapping ICD codes directly to GBD, we linked disease concepts through several classifications, including MeSH, ICD-10-CM, SNOMED CT, Disease Ontology, OMIM, and Orphanet, using shared concept identifiers where available. This allowed us to match disease concepts reported at different levels of specificity, without relying only on disease names or approximate text matching. Diseases were assigned at the publication (PMID) level, and a publication could be linked to more than one GBD cause when supported by its annotations. We excluded those ICD codes classified as non-specific or "garbage" categories in the GBD framework.

Using this approach, we were able to assign at least one valid GBD cause to 70.92% of RCTs-linked publications. Although higher coverage can be achieved using probabilistic or model-based methods [10], they introduce uncertainty that is difficult to verify and may affect how diseases are represented in the data. Because our study examines systematic inequalities between disease areas, we used this conservative approach to ensure that the mapping was interpretable, reproducible, and conceptually consistent.

**Inequality metrics**

We quantified research-disease inequality through the participation-to-burden ratio (PBR), calculated for each country-disease pair as:

$$PBR_{c,d,t} = \frac{\frac{P_{c,d,t}}{P_{g,d,t}}}{\frac{B_{c,d,t}}{B_{g,d,t}}}$$

where $P_{c,d,t}$ represents RCTs participants from country $c$ for disease $d$ in year $t$, $B_{c,d,t}$ represents DALYs for that country-disease-year, and $g$ denotes global totals. To address the right-skewed distribution of PBR values, we applied log-transformation (logPBR) for analyses. A logPBR of 0 indicates perfect proportionality: the country contributes to research in exact proportion to its disease burden. LogPBR > 0 indicates over-representation (the country contributes more research participation than its burden would predict), while logPBR < 0 indicates under-representation (the country contributes less participation than its burden warrants).

This metric is mathematically symmetric with respect to country and disease dimensions. PBR can be equivalently calculated as "country *c*'s share of participants in disease d relative to country *c*'s share of disease *d* burden" or as "disease *d*'s share of participants from country *c* relative to disease *d*'s share of burden in country *c*." This symmetry ensures that analytical choices do not bias results toward either the disease-centric or country-centric hypothesis. Secondary metrics, including Specialization Index, Gini Coefficient, Contribution to Inequality Score, are explained in **Supplementary Methods 3**.

**Inequality factor decomposition**

To separate the contributions of diseases and countries to total inequality, we used a complementary statistical approach that was less sensitive to how observations were grouped (**Supplementary Methods 3-5**).

*Bidirectional Theil Decomposition.* The Theil index [57], an entropy-based measure from information theory, decomposes total inequality into between-group and within-group components. We applied the Theil decomposition bidirectionally. First, grouping observations by disease yields:

$$T_{total} = T_{between-disease} + T_{within-disease}$$

where $T_{between-disease}$ captures inequality from differences in disease-average PBR values (whether some diseases are universally over- or under-researched), and $T_{within-disease}$ captures inequality from variation in PBR across countries within each disease (geographic disparities in participation for any given disease). Second, grouping observations by country yields:

$$T_{total} = T_{between-country} + T_{within-country}$$

where $T_{between-country}$ captures inequality from differences in country-average PBR values (whether some countries universally over- or under-contribute), and $T_{within-country}$ captures inequality from variation in PBR across diseases within each country (whether countries specialize in particular diseases). This approach tests whether conclusions depend on grouping choice (**Extended Data Fig. 4**). Country removal is used here as a sensitivity analysis.

*Two-Way ANOVA Variance Partitioning.* To estimate country and disease contributions simultaneously, we employed variance partitioning using a two-way fixed effects model:

$$PBR_{c,d,t} = \mu + \alpha_c + \beta_d + \gamma_t + \epsilon_{c,d,t}$$

where $c$ represents country fixed effects capturing time-invariant country characteristics, $d$ represents disease fixed effects capturing time-invariant disease characteristics, $t$ represents year fixed effects capturing temporal trends, and $c,d,t$ represents residual variation. We calculated the proportion of total variance explained by each component using partial $R^2$ values. This approach estimates country and disease contributions within the same model, allowing them to be compared directly.

*Hierarchical Variance Partitioning.* To quantify the contribution of country-level factors to explaining RCTs participation inequality, we employed hierarchical linear regression, using country-level PBR as the dependent variable (**Supplementary Methods 4**). For the full analysis, four categories were entered sequentially: Economic, Research Capacity, Health Infrastructure, and Governance and Societal Context. For each category, the incremental contribution was calculated as the additional variance explained by that category beyond all preceding categories in the sequence. For the policy-relevant residual analysis, we first regressed the PBR on the economic variables (GDP per capita, population size, and foreign aid received) and extracted residuals representing the unexplained inequality. We then regressed these residuals sequentially on Research Capacity, Health Infrastructure, and Governance and Societal Context.

*Shapley Value Decomposition.* To account for shared variance among factors and attribute explained variance to specific-country level factors without depending on their order of entry, we implemented Shapley value decomposition (**Supplementary Methods 4**). For each model, the Shapley value represents the average marginal contribution of each factor to the explained variance across different combinations of factors. Category-level contributions were obtained by summing the individual Shapley values for the factors within each category. We used 1,000 bootstrap iterations to estimate 95% confidence intervals for the percentage contribution of each category.

We employed variance partitioning and Shapley value decomposition to assess statistical association and explanatory power, specifically to account for multicollinearity among country-level factors. As several variables are strongly correlated with GDP and population size, isolating the independent effect of individual predictors is challenging. Given that the research ecosystem is complex and highly interconnected, our study focuses on macro-level determinants of RCTs participation rather than attempting to quantify the isolated impact of single factors on RCTs participation.

**Identification of limiting country-level factors**

For countries under-performing relative to predictions, we identified limiting research capacity, health infrastructure, and governance and societal factors through linear regression analysis (**Supplementary Methods 4**). We estimated disease-specific regression models predicting logPBR from country-level factors, yielding predicted values and residuals:

$$Residual_{i,d} = log\ log\ (PBR)_{i,d} - log\ log\ (\hat{P}BR)_{i,d}$$

To account for variance differences across disease categories, country-disease pairs were classified into performance groups based on disease-specific residual standard deviations (SD):

- Over-performing (Residual>0.9*SD): RCTs participation significantly exceeds structural predictions.
- Under-performing (Residual<-0.5*SD): RCTs participation falls significantly below structural predictions.
- As-expected (|Residual| ≤ 0.1*SD): RCTs participation closely matches structural predictions.
- Intermediate (-0.5*SD ≤ Residual ≤ 0.9*SD, excluding |Residual| ≤ 0.1*SD): Country-disease pairs with residuals between the expected and over- or under-performing thresholds. Because these cases did not clearly indicate over- or under-performance, we excluded them from the analysis

For under-performing observations, we identified limiting factors by examining factor coefficients within conceptual categories. For each category $b$, we computed average absolute coefficient magnitude among significant factor ($p < 0.1$):

$$\overline{\beta_b} = \frac{1}{n}\sum_{j \in b} |\beta_j| \cdot 1(p_j < 0.1)$$

The category with largest $\overline{\beta_b}$ was classified as the primary limiting factor, indicating which policy lever would have the greatest marginal impact. Country-disease pairs with multiple categories showing comparably strong effects ($\overline{\beta_b} > 0.7 \cdot \max(\overline{\beta_b})$) were classified as requiring "Multiple Factors" coordinated intervention.

To assign country-level factors to over-performing combinations without modifying the regression model, we applied a hierarchical matching procedure. Factor labels were inferred by matching over-performing combinations to dominant limiting-factor patterns observed among under-performing combinations at the disease-level, then country-level, with a global fallback where necessary.

For as-expected combinations, factor labels reflect the dominant contributor to expected performance. Each combination was assigned to the factor corresponding to the largest normalized component of authorship, disease burden, or participant enrollment, representing the primary alignment underlying its expected research-burden relationship.

**Intervention scenarios**

To quantify how reducing structural misalignment might impact global research inequality, we simulated two intervention scenarios at the country-level PBR:

1. Full Alignment: All 180 countries gradually shift their PBR values toward the global median (0.249, **Supplementary Data**). This scenario represents the theoretical maximum reduction in inequality achievable through complete structural harmonization.
2. Targeted Alignment: Only the most misaligned countries-those with PBR values deviating most from the median-are incrementally adjusted. This scenario reflects a resource-efficient strategy prioritizing high-impact interventions.

For each scenario, we calculated the national PBR's Gini coefficient reduction using bootstrap resampling (200 iterations) across four intervention intensities: adjusting 25%, 50%, 75%, and 100% of countries for Full Alignment, and 10%, 20%, 30%, and 40% for Targeted Alignment. Percentage reduction was computed as $100 \times (G_{baseline} - G_{adjusted})/G_{baseline}$, where $G_{baseline}=0.772$ represents the observed inequality.

To compare strategy efficiency, we calculated reduction per country adjusted: Efficiency=Percentage reduction/Percentage of countries adjusted. Statistical significance was assessed through paired t-tests comparing baseline versus adjusted Gini distributions across bootstrap samples.

**Global research network features after interventions**

To understand how these two realignments might reshape global research collaboration patterns, we constructed a country-factor-disease network and simulated its evolution under intervention scenarios.

The network comprised 450 nodes representing unique country-factor combinations (e.g., "USA-Research_Capacity") and 25,047 edges connecting nodes that participate in the same disease. Node attributes included: factor (Research_Capacity, Health_Infrastructure, Governance, Multiple_Factors), performance status (Over_performing, As_Expected, Under_performing), and disease count (node size). Edge weights reflected the number of shared diseases between connected nodes. We analyzed four network measures that capture different dimensions of collaboration structure:

1. Network Density [58]: Proportion of possible connections that exist, measuring overall connectivity.
2. Homophily [59]: Proportion of edges connecting nodes with the same factor, measuring segregation by factor.
3. Modularity [60]: Strength of community structure, calculated using the Louvain algorithm.
4. Average Path Length [58]: Mean shortest distance between all node pairs, measuring network integration.

We modeled network evolution by linking changes in these metrics to reductions in structural inequality (Gini coefficient). Uncertainty was estimated through parametric bootstrap resampling (100 iterations), with confidence intervals reflecting both measurement uncertainty in baseline metrics and variability in sensitivity relationships.

## Data availability

The data assembled for this study are available and can be accessed at https://doi.org/10.5281/zenodo.20695734. Source data are provided with this paper.

## Code availability

The computer code used to perform the analyses in this study is available and can be accessed via the following link: https://doi.org/10.5281/zenodo.21593311.

## References

[54]Institute for Health Metrics and Evaluation (IHME) Global Burden of Disease Study 2023 (GBD 2023) Data Resources (IHME, 2024).
[55]Zotova, E., Cuadros, M., & Rigau, G. (2022, June). ClinIDMap: Towards a clinical IDs mapping for data interoperability. In *Proceedings of the Thirteenth Language Resources and Evaluation Conference* (pp. 3661-3669).
[56]Del Valle, E. P. G., García, G. L., Santamaría, L. P., Zanin, M., Ruiz, E. M., & Rodríguez-González, A. (2021). DisMaNET: A network-based tool to cross map disease vocabularies. *Computer Methods and Programs in Biomedicine*, 207, 106233.
[57]Theil, H. (1967) *Economics and Information Theory*. North-Holland Publishing Company, Amsterdam.
[58]Wasserman, S. (1994). *Social network analysis: Methods and applications*.Cambridge University Press.
[59]McPherson, M., Smith-Lovin, L., & Cook, J. M. (2001). Birds of a feather: Homophily in social networks. *Annual Review of Sociology*, 27(1), 415-444.
[60]Newman, M. E., & Girvan, M. (2004). Finding and evaluating community structure in networks. *Physical Review E*, 69(2), 026113.

## Acknowledgements

We thank M. Rousseau, J. Du, and P. Xiao for their valuable assistance in rationalization. We also thank D. Kozlowski, PJ. Chen, LF. Wu, YF. Duan, QP. Wang, and GX. He, Y. Xia for their excellent research support. J.H. thanks Google GCP Gemini Academic Program award for providing credits on Gemini API. W.L. acknowledges funding from the 2025 Major Special Project of the Ministry of Education on Philosophy and Social Sciences Research (Grant Number: 2025JZDZ101). V.L. acknowledges funding from the Social Science and Humanities Research Council of Canada Pan-Canadian Knowledge Access Initiative Grant (Grants 1007-2023-0001 and 435-2025-1001), and the Fonds de recherche du Québec-Société et Culture through the Programme d'appui aux Chaires UNESCO (Grant Number 338828).

## Author contributions

W.L., J.H., and V.L. devised the original idea. W.L. and A.A.D. conceptualized the study, drafted, and wrote the final manuscript. W.L. led the study and analyses. W.L., J.H., Z.L., and V.L. assembled and analyzed the data. W. L. provided access to data sources, methods, results, and interpretations. All authors edited the manuscript.

## Competing interests

The authors declare no competing interests.

## Additional information

Supplementary information is available for this paper.

# Extended Data

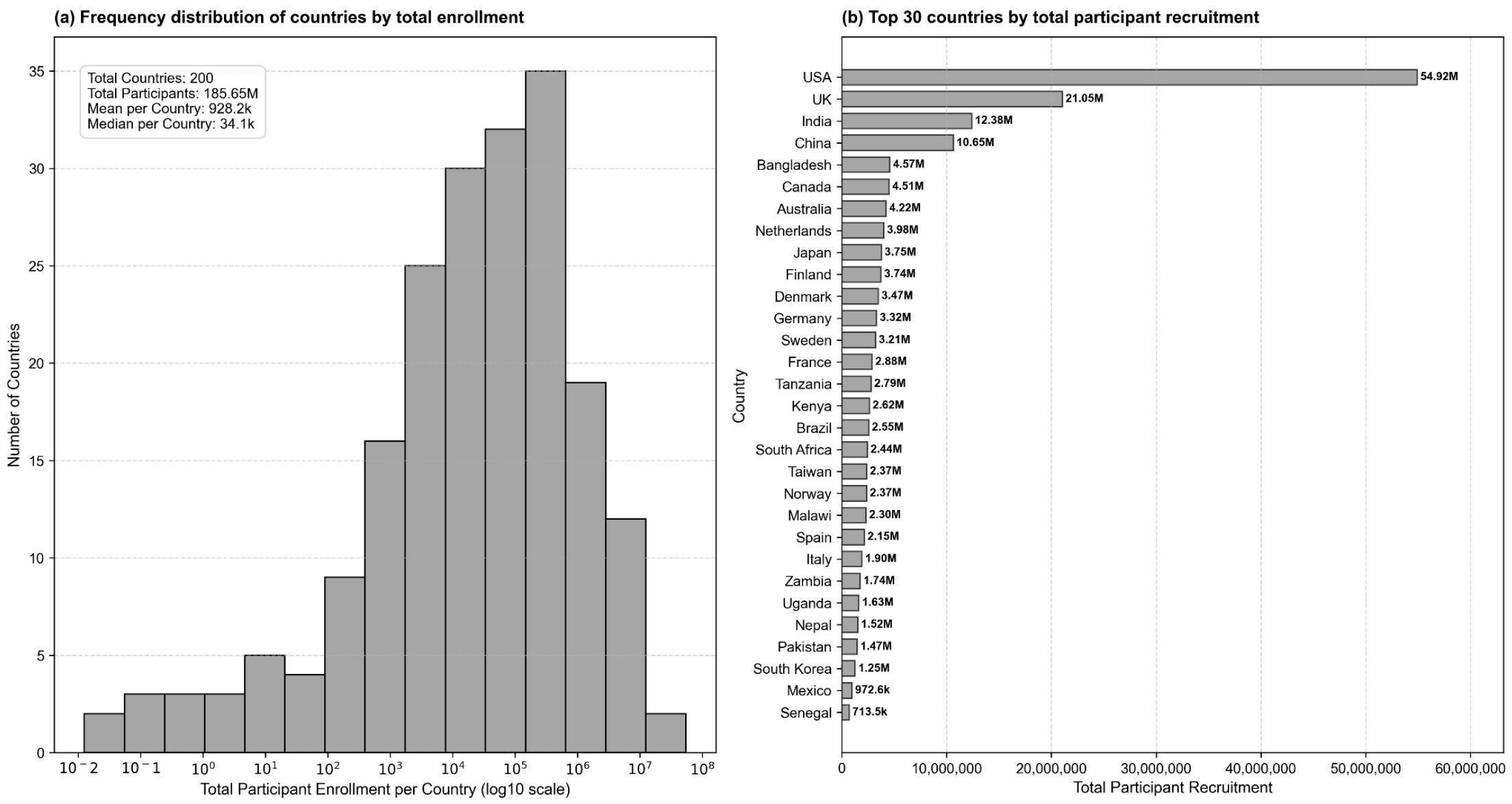

**Extended Data Fig. 1| Geographic distribution of RCTs participants by country (n=200).** Participant totals represent cumulative enrollment across all studies conducted in each country from 2000-2024. Panel a, Histogram showing the frequency distribution of countries by total participant enrollment in the countries (plotted on a log10 scale, with a summary stats inset displaying total countries, total recruitment, mean, and median enrollment values). Panel b, Horizontal bar chart showing the top 30 countries by total participant recruitment across all studies in the dataset. Countries are ordered by absolute participant numbers, with numerical annotations indicating total recruitment.

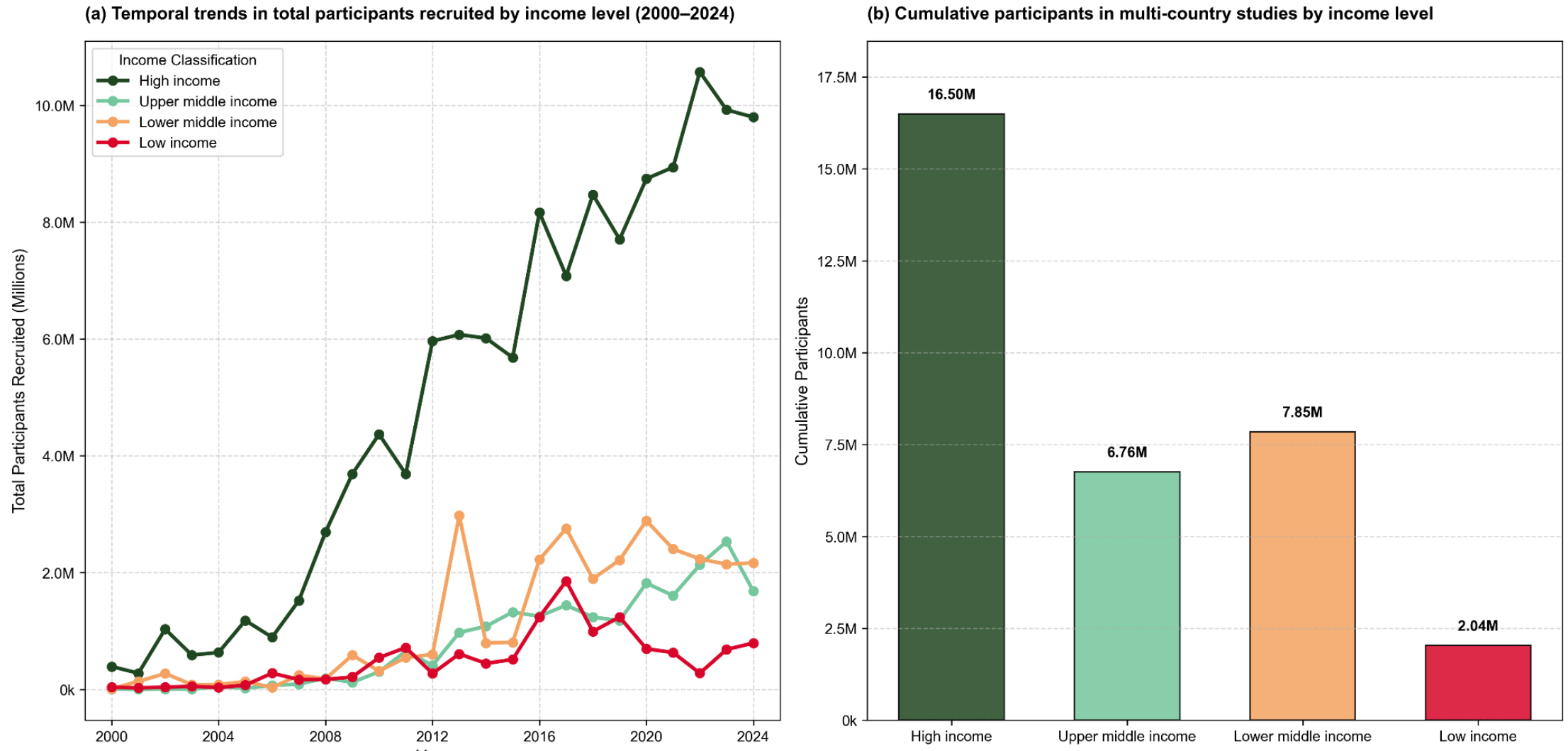

**Extended Data Fig. 2| RCTs participant recruitment patterns by World Bank income classification (n=200).** Panel a shows temporal trends in total participants recruited by income level from 2000-2024, with high-income countries (dark green) demonstrating exponential growth reaching 10.5M participants by 2022, while other income groups show modest, stable recruitment levels. Panel b displays cumulative participants in multi-country studies by income level, revealing extreme inequality with high-income countries recruiting 16.50M participants compared to 2.04M in low-income countries.

Comprehensive statistical testing reveals significant associations between country income classification and research participation rates. Kruskal-Wallis test confirms significant differences across income groups ($H = 29.871$, $p = 1.4692e-06$), while Spearman correlation demonstrates slightly positive association between income level and participation rate ($\rho = 0.233$, $p = 9.0371e-05$). High-income countries average 74,654.0 participants per million population compared to 8814.0 in lower-middle-income countries, despite low-income countries showing elevated rates (14,215.3 per million) due to specific high-participation outliers. Chi-square analysis confirms significant association between recruitment patterns and income classification ($\chi^2 = 401.349$, $p = 1.1296e-86$), indicating systematic economic association with global research participation access.

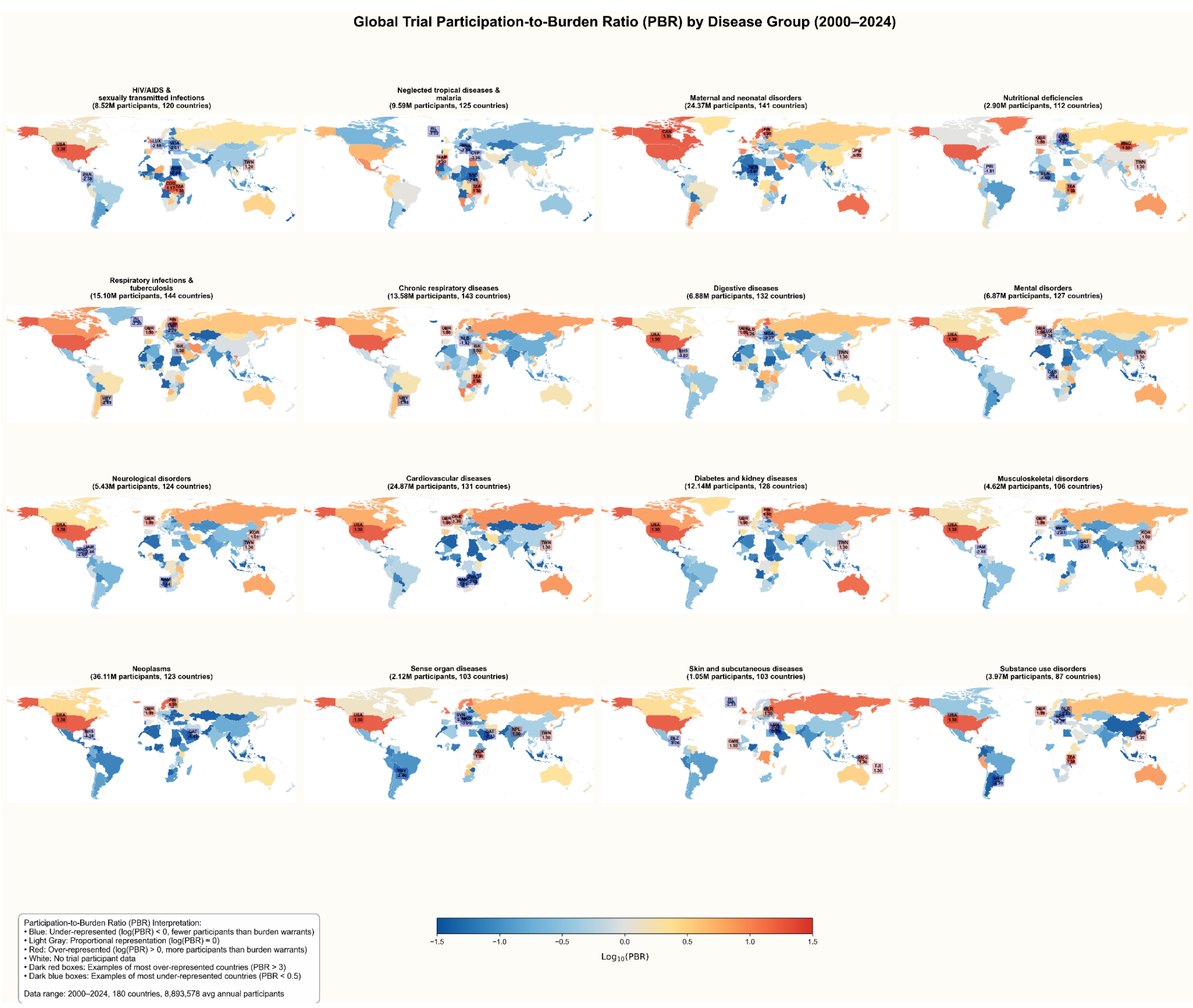

**Extended Data Fig. 3|Global RCTs participation-to-burden ratio (PBR) by Level 2 disease categories (2000-2024) (n=180).** Sixteen-panel world map displaying geographic patterns of research participation relative to disease burden for each custom disease category. Colors represent $\log_{10}$-transformed PBR values using a diverging colormap: blue indicates under-representation (log(PBR) < 0, fewer participants than burden share warrants), light gray indicates proportional representation (log(PBR) ≈ 0), and red indicates over-representation (log(PBR) > 0, more participants than burden share). White areas indicate no participant data. Each panel title shows total participants and number of countries with data. The color bar is centered at 0 (corresponding to PBR = 1) with range from -1.5 to +1.5 on $\log_{10}$ scale.

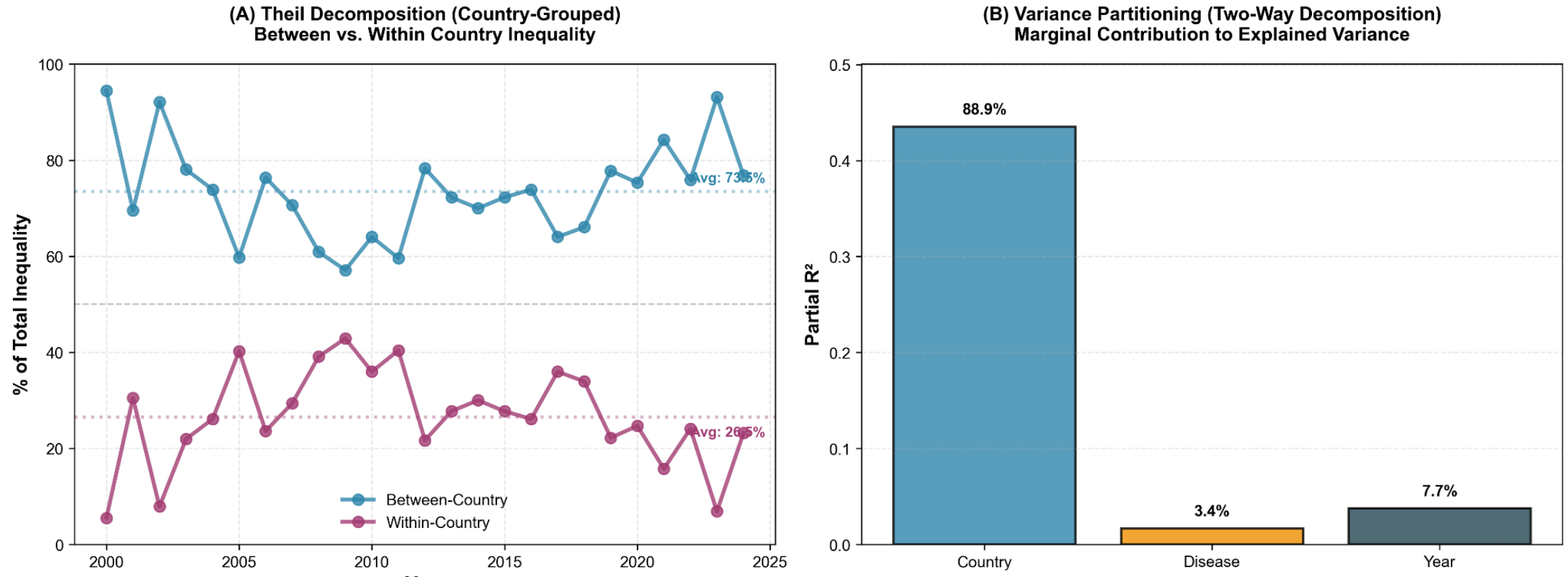

**Extended Data Fig. 4| Symmetric decomposition analyses quantifying relative contributions of country versus disease factors.** Two-panel analysis providing complementary perspectives on inequality drivers: Panel A shows Theil decomposition grouped by country, partitioning total inequality into between-country variance versus within-country variance across years 2000-2024, with a 50% reference line indicating equal contribution. Panel B presents two-way variance partitioning using fixed effects regression, calculating partial $R^2$ values representing the marginal contribution of country, disease, and year fixed effects to explained variance in participation-to-burden ratios. Percentage labels indicate each component's share of total explained variance.

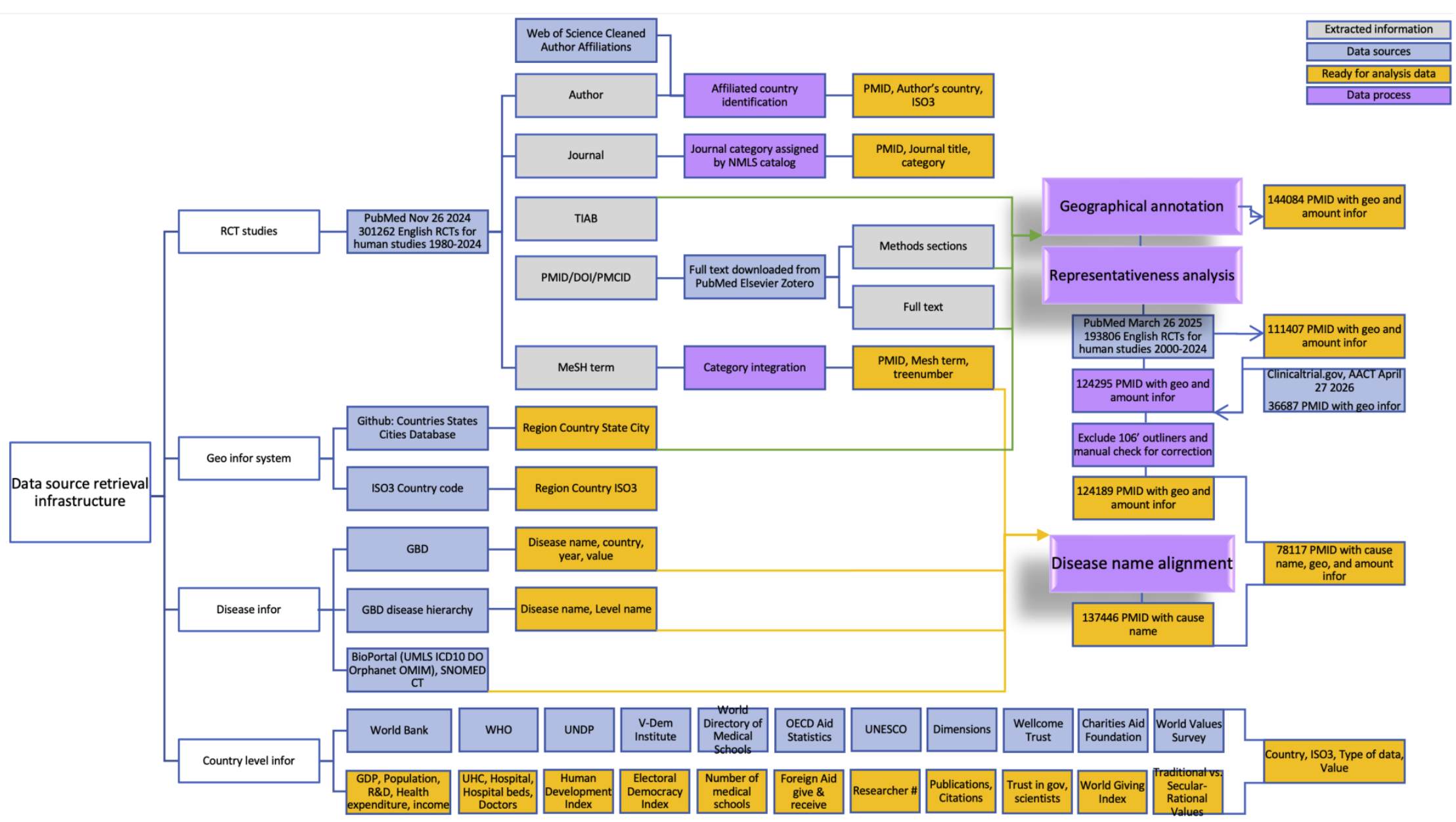

**Extended Data Fig. 5| Study design and data integration workflow.** Schematic overview of data sources, processing pipeline, and analytical framework for examining global disparities in RCTs participation. The study integrated five primary data sources: PubMed bibliographic records (n=301,262 studies, 1980-2024), Global Burden of Disease estimates (939K country-year-cause observations), author affiliation data, participant geographic and demographic information extracted via AI-assisted methods, and country level indicators from international databases. After temporal restriction to 2000-2024 and sequential filtering for geographic annotation and disease mapping, and complementing with Clinicaltrials.gov data, the final analytical datasets comprised 78,117 studies with complete participant and disease information. PMID, PubMed identifier; GBD, Global Burden of Disease; AI, artificial intelligence; TIAB, title and abstract; BioPortal's datasets see Supplementary Table 5.

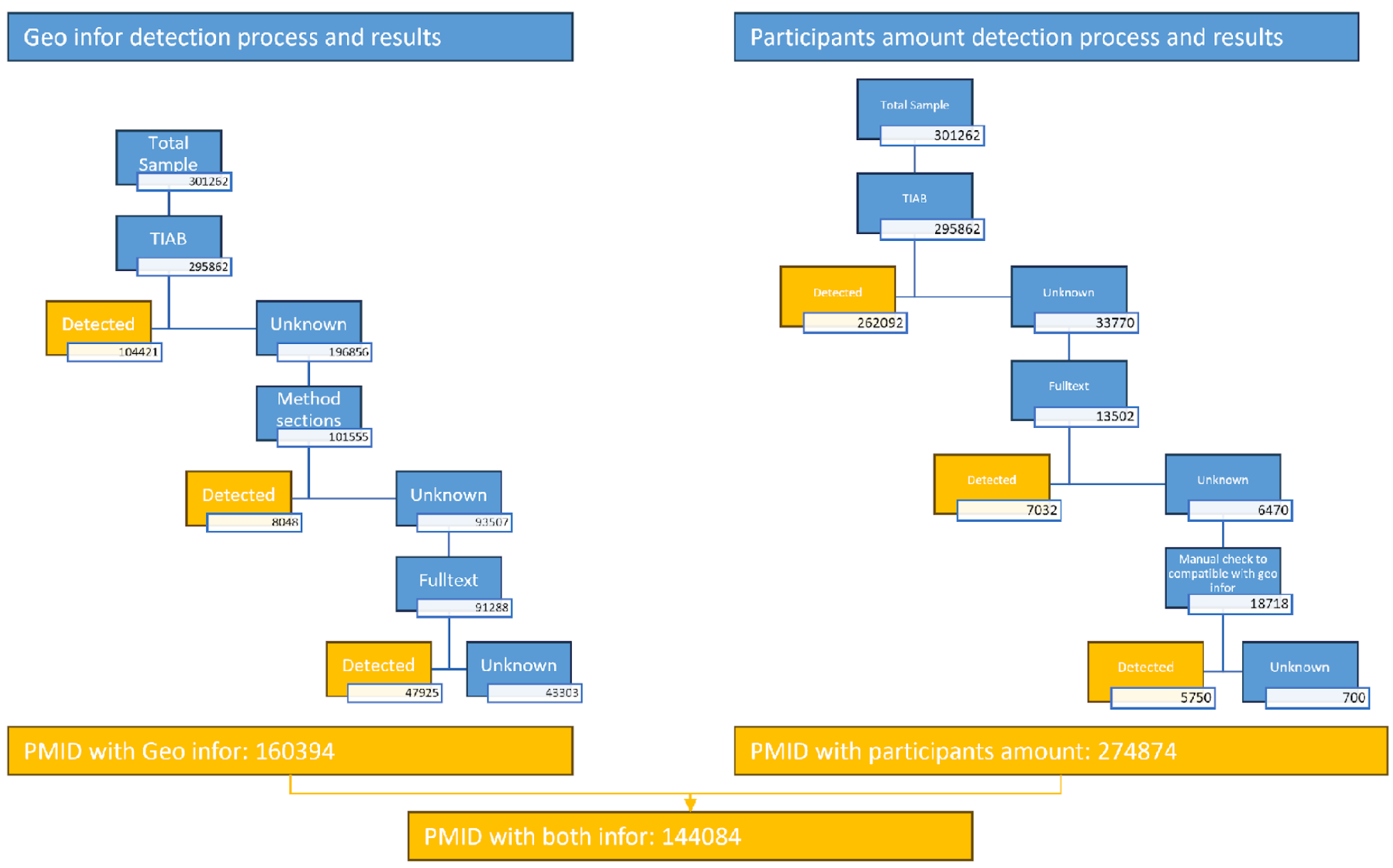

**Extended Data Fig. 6| AI-assisted geographic annotation and participant extraction pipeline for first round**. Multi-stage workflow for extracting participant geography and enrollment numbers from RCTs publications. Text inputs were sourced hierarchically from abstracts, methods sections, and full-text articles. Geographic locations were identified using AI-assisted named entity recognition targeting universities, hospitals, cities, and countries. Participant counts were extracted using rule-based patterns and validated through manual review. Methodological details of AI-assisted extraction results see Supplementary Table 3.

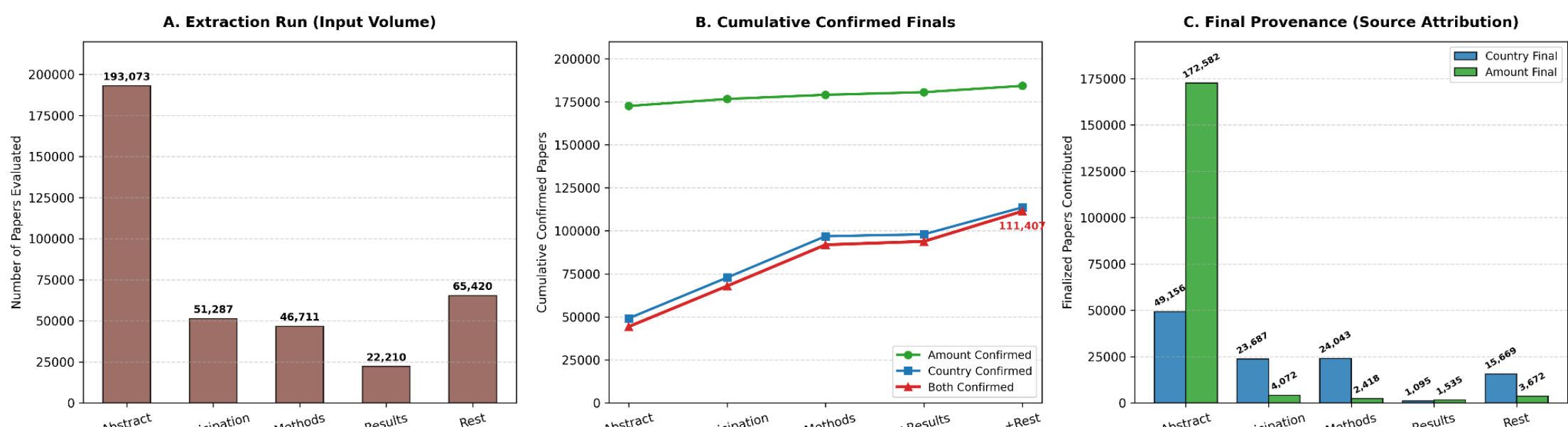

**Extended Data Fig. 7| Performance and attribution of the multi-stage extraction cascade for the second round.** Three-panel evaluation quantifying pipeline performance across 193,806 total papers. Panel A shows total extraction run volume evaluating paper instances per section. Panel B tracks cumulative yield of confirmed metadata through the sequential cascade (Title and Abstract + Participation + Methods +Results + Rest), both fields resolved to 111,407. Panel C presents final source attribution for resolved database records (country_final and amount_final). More details are shown in Supplementary Table 4.

# Supplementary Tables

**Supplementary Table 1. Temporal distribution and sampling strategy for AI model development**. Annual distribution of RCTs publications used for training and validating geographic annotation and participant extraction models (1980-2024). A stratified sampling approach was employed to ensure representative coverage across publication years, with sampling proportions adjusted for annual publication volumes. The validation dataset comprised 359 articles selected proportionally from 301,262 total publications for the first round of AI-assisted information extraction process in 2024. Due to the suboptimal precision and recall, we conducted a second round of sampling and information extraction process to improve them with updated LLMs. Second round of the sampling dataset comprised 1,482 articles selected proportionally from 193,806 total publications between 2000-2024.

| Year | Count | Sampling for first round | Sampling for second round |
| --- | --- | --- | --- |
| 2024 | 12862 | 13 | 99 |
| 2023 | 16867 | 19 | 102 |
| 2022 | 20439 | 20 | 113 |
| 2021 | 21247 | 22 | 105 |
| 2020 | 19328 | 20 | 93 |
| 2019 | 18446 | 19 | 86 |
| 2018 | 17814 | 18 | 83 |
| 2017 | 17658 | 18 | 87 |
| 2016 | 17525 | 19 | 84 |
| 2015 | 18097 | 18 | 84 |
| 2014 | 18084 | 19 | 82 |
| 2013 | 17132 | 18 | 75 |
| 2012 | 14707 | 15 | 60 |
| 2011 | 12716 | 14 | 56 |
| 2010 | 10854 | 10 | 48 |
| 2009 | 9421 | 9 | 41 |
| 2008 | 8062 | 9 | 35 |
| 2007 | 6879 | 7 | 28 |
| 2006 | 6264 | 7 | 24 |
| 2005 | 5896 | 6 | 21 |
| 2004 | 5067 | 7 | 20 |
| 2003 | 4260 | 3 | 17 |
| 2002 | 3556 | 4 | 14 |
| 2001 | 3065 | 4 | 12 |
| 2000 | 3329 | 4 | 13 |
| 1999 | 3362 | 4 | |
| 1998 | 3162 | 4 | |
| 1997 | 2901 | 3 | |
| 1996 | 2425 | 3 | |
| 1995 | 2334 | 3 | |
| 1994 | 1940 | 2 | |
| 1993 | 1424 | 2 | |
| 1992 | 1309 | 2 | |
| 1991 | 1322 | 2 | |
| 1990 | 1260 | 2 | |
| 1989 | 965 | 1 | |
| 1988 | 806 | 1 | |
| 1987 | 794 | 1 | |
| 1986 | 748 | 1 | |
| 1985 | 652 | 1 | |
| 1984 | 548 | 1 | |
| 1983 | 613 | 1 | |
| 1982 | 560 | 1 | |
| 1981 | 453 | 1 | |
| 1980 | 486 | 1 | |
| Total | 337,639 (distinct 301,262) | 359 | 1482 |

**Supplementary Table 2. Performance evaluation of nine prompt strategies for geographic annotation and participant count extraction.** Each strategy was evaluated on a stratified sample of 359 RCTs articles using precision, recall, and F1-score. Geographic annotation (Geo) and participant count (#) extractions were evaluated separately to optimize task-specific performance. Strategy selection balanced for both Geo and # tests. Therefore, Strategy 8 and 9 were chosen due to the optimal performance (F1=0.879 and 0.855).

| Strategy | Precision | Recall | F1 |
|---|---|---|---|
| (1) Least guidance | 0.651 | 0.736 | 0.691 |
| (2) Human brain guidance without examples | 0.814 | 0.814 | 0.814 |
| (3) Human brain guidance with examples | 0.869 | 0.822 | 0.845 |
| (4) Geo and # in order combined with examples (for Geo) | 0.833 | 0.271 | 0.409 |
| (5) Geo and # in order combined with examples (for #) | 0.972 | 0.795 | 0.875 |
| (6) Geo and # in reversed order combined with examples (for Geo) | 0.821 | 0.248 | 0.381 |
| (7) Geo and # in reversed order combined with examples (for #) | 0.958 | 0.808 | 0.877 |
| (8) Geo and # separately with examples (for Geo) | 0.865 | 0.845 | 0.855 |
| (9) Geo and # separately with examples (for #) | 0.954 | 0.815 | 0.879 |

**Supplementary Table 3. Performance comparison of geographical annotation methods for the first round.** Four approaches were evaluated for extracting enrollment location and participant size from RCTs publications using a validation dataset of 359 publications. Manual annotation by domain experts served as the gold standard. String matching provided rapid but limited coverage through exact text matching of country names. Machine learning was employed named entity recognition with location-specific training data. LLaMA and Gemma2 were a general-purpose open-source model.

| Task | Model | Approach | Precision | Recall | F1 Score |
|---|---|---|---|---|---|
| Geographical annotation | Model 1 | String match | 0.989 | 0.674 | 0.989 |
| | Model 2 | Machine learning | 0.372 | 0.450 | 0.372 |
| | Model 3 | LLaMA | 0.674 | 0.744 | 0.674 |
| | Model 4 | Gemma2 | 0.865 | 0.845 | 0.865 |
| Participant size | Model 3 | LLaMA | 0.952 | 0.769 | 0.952 |
| | Model 4 | Gemma2 | 0.954 | 0.815 | 0.954 |

**Supplementary Table 4. Performance comparison of geographical annotation methods for the second round.**

Panel A: Three methods were evaluated for extracting enrollment location and A. Full text, untrimmed: The entire Markdown-rendered full text (title, abstract, body, author affiliations, references, acknowledgements) is passed to the model in a single call. B. Full text, trimmed: Identical to Method A except that the Markdown is first passed through a rule-based shrinking step that removes content unlikely to describe the study sample, including: authors, affiliations, discussion, conclusions, and references. C. Section cascade: The Markdown full text is split by headings, and each section is assigned by rule-based title matching to one of four buckets, ordered by their prior likelihood of stating the enrollment country and size. Acc. is short for accuracy. Mean and Median represent the mean and median numbers of input tokens per paper.

| **Method** | **Country Precision** | **Country Recall** | **Country F1** | **Country Acc.** | **Amount Acc.** | **Amount F1** | **Mean** | **Median** |
|---|---|---|---|---|---|---|---|---|
| A. Full text, untrimmed | 0.939 | 0.956 | 0.948 | 0.944 | 0.965 | 0.969 | 22,304 | 22,401 |
| B. Full text, trimmed | 0.965 | 0.952 | 0.959 | 0.957 | 0.967 | 0.971 | 7,245 | 6,845 |
| C. Section cascade | 0.971 | 0.939 | 0.955 | 0.962 | 0.967 | 0.971 | 4,299 | 2,866 |

Panel B: Subgroup performance metrics for AI extraction steps across publication epochs, geographic regions, and study scope.

| **Subgroups** | | **Sample Size (n)** | **Country Precision** | **Country Recall** | **Country F1** | **Country Acc.** | **Amount Acc.** |
|---|---|---|---|---|---|---|---|
| Publication Epoch | 2000–2009 (Early) | 225 | 0.938 | 0.889 | 0.913 | 0.924 | 0.987 |
| | 2010–2017 (Middle) | 576 | 0.984 | 0.941 | 0.962 | 0.976 | 0.957 |
| | 2018–2024 (Recent) | 681 | 0.974 | 0.964 | 0.969 | 0.966 | 0.966 |
| Geographic Region | Europe | 350 | 0.980 | 0.980 | 0.980 | 0.980 | 0.969 |
| | North America | 341 | 0.977 | 0.977 | 0.977 | 0.977 | 0.962 |
| | Asia | 265 | 0.974 | 0.974 | 0.974 | 0.974 | 0.970 |
| | Africa | 44 | 1.000 | 1.000 | 1.000 | 1.000 | 0.932 |
| | South America | 33 | 1.000 | 1.000 | 1.000 | 1.000 | 1.000 |
| | Oceania | 33 | 0.970 | 0.970 | 0.970 | 0.970 | 0.970 |
| Study Scope | Single-Country Studies | 1,066 | 0.978 | 0.978 | 0.978 | 0.978 | 0.966 |
| | Multi-Country Studies | 96 | 0.993 | 0.872 | 0.929 | 0.885 | 0.979 |
| No Location Stated | | 320 | 0.926 | 0.938 | 0.932 | 0.938 | 0.959 |

**Supplementary Table 5. Multi-pathway mapping framework for linking MeSH terms to GBD disease categories.** RCTs MeSH descriptors were systematically mapped to Global Burden of Disease cause names through six intermediate ontological systems. Direct mappings were prioritized where available, followed by multi-step bridging via concept unique identifiers (CUIs). MeSH tree numbers (n=64,883) served as input, with intermediate systems including Disease Ontology (DO), Orphanet Rare Disease Ontology (ORDO), Online Mendelian Inheritance in Man (OMIM), and Systematized Nomenclature of Medicine Clinical Terms (SNOMED). The framework achieved 6,595 unique mapping pathways, successfully linking 137,446 studies (70.92% of 193,806 post-2000 publications) to 296 distinct GBD cause names. Final coverage included 296 of 308 ICD-10 mapped cause names (96.10%), demonstrating comprehensive disease representation across RCTs literature. UMLS, Unified Medical Language System; UID, unique identifier.

| **RCTs items** | **Ontologies** | | | **GBD cause names can be mapped to** | **# mapped unique UID** |
|---|---|---|---|---|---|
| MeSH tree number | UMLSUID | | | ICD10 | 2,141 |
| | DO | MeSHCUI | | | 1,325 |
| | | UMLSUID | | | 1,242 |
| | | direct | | | 1,271 |
| | ORDO | MeSHCUI | | | 531 |
| | | UMLSUID | | | 538 |
| | | Direct | | | 625 |
| | OMIM | UMLSUID | | | 1,158 |
| | | ORDO | UMLSUID | | 273 |
| | SNOMED | ORDO | | | 601 |
| | | ORDO | UMLSUID | | 731 |
| | | UMLSUID | | | 1,975 |
| | | DO | | | 977 |
| | | DO | UMLSUID | | 974 |
| | Direct | | | | 2,228 |
| Total distinct mapping | | | | | 6,595 |
| Total distinct mapped PMID in 193,806 studies | | | | | 137,446(70.92%) |
| Total mapped cause names | | | | | 296 |

| Total mapped cause names in 308 cause names with ICD10 | 296(96.10%) |
|---|---|

*Note: Multiple mapping pathways for individual disease concepts reflect the redundancy built into the framework to maximize coverage while maintaining precision.*

## Supplementary Table 6. GBD Level 2 Disease Inequality Analysis

| Rank | Disease Category | Total Participants | Global North Participant Share | Global North DALY Share (Burden) | Global South DALY Share (Burden) | Inequality Ratio (North Part% / North DALY%) |
|---|---|---|---|---|---|---|
| 1 | Maternal and neonatal disorders | 24.33M | 59.00% | 1.65% | 98.35% | 35.69x |
| 2 | Respiratory infections and tuberculosis | 15.05M | 50.85% | 3.84% | 96.16% | 13.24x |
| 3 | Neglected tropical diseases and malaria | 9.59M | 2.89% | 0.23% | 99.77% | 12.46x |
| 4 | Nutritional deficiencies | 2.90M | 25.21% | 2.12% | 97.88% | 11.90x |
| 5 | HIV/AIDS and sexually transmitted infections | 8.52M | 15.81% | 1.36% | 98.64% | 11.58x |
| 6 | Cardiovascular diseases | 24.86M | 71.87% | 12.70% | 87.30% | 5.66x |
| 7 | Digestive diseases | 6.87M | 60.89% | 12.58% | 87.42% | 4.84x |
| 8 | Diabetes and kidney diseases | 12.13M | 75.96% | 15.87% | 84.13% | 4.79x |
| 9 | Mental disorders | 6.81M | 81.69% | 17.74% | 82.26% | 4.61x |
| 10 | Chronic respiratory diseases | 13.53M | 56.82% | 13.70% | 86.30% | 4.15x |
| 11 | Skin and subcutaneous diseases | 1.05M | 63.56% | 16.07% | 83.93% | 3.95x |
| 12 | Musculoskeletal disorders | 4.62M | 86.64% | 22.63% | 77.37% | 3.83x |
| 13 | Neoplasms (Cancer) | 36.10M | 88.36% | 23.53% | 76.47% | 3.75x |
| 14 | Neurological disorders | 5.42M | 85.43% | 23.17% | 76.83% | 3.69x |
| 15 | Sense organ diseases | 2.11M | 45.29% | 13.89% | 86.11% | 3.26x |
| 16 | Substance use disorders | 3.96M | 92.26% | 29.52% | 70.48% | 3.13x |

*Note: Calculated on the disease-matched country dataset across 16 GBD Level 2 disease categories (2000–2024). Global North is defined as High-Income countries.*

**Supplementary Table 7. Data sources, coverage, and descriptive statistics for country-level indicators**

| **Category** | **Function** | **Variables** | **Indicators** | **Country coverage** | **Part 1*** | **Part 2*** |
|---|---|---|---|---|---|---|
| Disease Burden | Demand for research | DALYs | Number of DALYs in each country each year | 201 | Yes. Dependent Variable | Yes. Use residual |
| Economic | Country economic conditions | Wealth | GDP per capita | 190 | Yes. Log | No. |
| | | Scale | Population | 204 | Yes. Log | No. |
| | | External source | Foreign aid received as a share of national income | 142 | Yes. Raw. | No. |
| | | | Foreign aid given as a share of national income | 49 | No. Insufficient data | No. |
| Research Capacity | Can this country produce science | Invest in research | R&D expenditure as a share of GDP | 150 | Yes. Raw | Yes. Raw. |
| | | Research force | Researchers per million | 142 | Yes. Log. | Yes. Log. |
| | | Research output | Total publications | 180 | Yes. Log. | Yes. Log. |
| | | Research impact | Total citations | 180 | Yes. Log. | Yes. Log. |
| Health Infrastructure | Can this country recruit and manage patients | Invest in health | Health expenditure per capita | 178 | Yes. Log. | Yes. Log. |
| | | Health potential | Medical schools-2024 | 165 | Yes. Log. | Yes. Log. |
| | | Clinical capacity | Hospital-2014 | 41 | No. Insufficient data | No. |
| | | | Hospital beds per capita | 183 | No. Explanatory power trade off. | No. |
| | | | Doctors per 10,000 | 186 | No. Explanatory power trade off. | No. |
| | | Basic supplies | Universal health coverage (UHC) index | 202 | Yes. Raw | Yes. Raw |
| | | | Sanitation as a share of population | 202 | Yes. Raw | Yes. Raw |
| Governance and societal context | The country's enabling environment | Institutional quality | Electoral Democracy Index | 174 | Yes. Raw. | Yes. Raw. |
| | | Social trust | Trust in Government Index-2020 | 107 | No. Insufficient data. | No. |
| | | | Trust in Scientists Index-2018 | 113 | Yes. Raw. | Yes. Raw. |
| | | Development | HDI | 184 | Yes. Raw. | Yes. Raw. |

| | | | | |
|---|---|---|---|---|
| Altruism | World Giving Index | 163 | Yes. Raw. | Yes. Raw. |
| Cultural values | Traditional/Secular Rational Values-tradrat5 | 95 | No. Insufficient data. | No. |

*Note: Part 1 and 2 show if the variables are selected, the reasons for selection, and the transformation of the variable.*

**Supplementary Table 8. Hierarchical variance partitioning and Shapley decomposition results of full analysis model**

| Category / Variable | HVP R2 | HVP % | Shapley R2 | Shapley % | Explanatory Range |
|---|---|---|---|---|---|
| **Economic (Total)** | **0.261** | **51.86 %** | **0.260** | **51.65%** | 51.65% – 58.96% |
| GDP per capita | 0.0200 | 3.97% | 0.0395 | 7.84% | 3.97% – 12.80% |
| Population | 0.1533 | 30.45% | 0.1615 | 32.09% | 25.10% – 32.09% |
| Foreign aid received | 0.0878 | 17.44% | 0.0590 | 11.72% | 11.72% – 17.44% |
| **Research Capacity (Total)** | **0.125** | **24.77 %** | **0.155** | **30.79%** | 14.37% – 61.92% |
| R&D expenditure | 0.0909 | 18.05% | 0.0644 | 12.78% | 12.78% – 31.84% |
| Publications | 0.0014 | 0.27% | 0.0378 | 7.52% | 0.27% – 9.42% |
| Citations | 0.0296 | 5.88% | 0.0464 | 9.22% | 5.88% – 13.28% |
| Researchers per million | 0.0029 | 0.57% | 0.0064 | 1.27% | 0.57% – 16.03% |
| **Health Infrastructure (Total)** | **0.092** | **18.30 %** | **0.066** | **13.07%** | 13.07% – 52.66% |
| Health exp. per capita | 0.0267 | 5.29% | 0.0217 | 4.30% | 4.30% – 20.62% |
| UHC index | 0.0050 | 0.99% | 0.0067 | 1.34% | 5.20% – 47.84% |
| Medical schools | 0.0572 | 11.36% | 0.0262 | 5.20% | 0.99% – 4.26% |
| Sanitation | 0.0033 | 0.65% | 0.0112 | 2.22% | 0.65% – 4.15% |
| **Governance and Societal Context (Total)** | **0.026** | **5.07%** | **0.023** | **4.49%** | 1.00% – 13.06% |
| HDI | 0.0025 | 0.50% | 0.0034 | 0.67% | 0.50% – 1.65% |
| Democracy index | 0.0001 | 0.02% | 0.0036 | 0.72% | 0.02% – 2.80% |
| Altruism | 0.0195 | 3.88% | 0.0137 | 2.73% | 2.73% – 7.64% |
| Trust in scientists | 0.0034 | 0.67% | 0.0019 | 0.37% | 0.30% – 0.67% |
| **Total Model** | **0.503** | **100.00%** | **0.503** | **100.00%** | 48.5% – 50.3% |

*Note: Detailed Level 2 variable decomposition for the Full 15-variable structural model (evaluated on the unified 180 countries). HVP R2 measures incremental variance explained sequentially following structural developmental ordering; Shapley R2 evaluates independent contributions averaged over all variable permutations. Explanatory range is the explanatory value ranges across model specifications. The unexplained is the opposite proportion of 1.*

**Supplementary Table 9. Hierarchical variance partitioning and Shapley decomposition results of policy relevant residual model**

| Category / Variable | HVP R2 | HVP % | Shapley R2 | Shapley % |
|---|---|---|---|---|
| **Research Capacity (Total)** | **0.154** | **61.92%** | **0.137** | **55.23%** |
| R&D expenditure | 0.0790 | 31.84% | 0.0749 | 30.17% |
| Publications | 0.0108 | 4.36% | 0.0234 | 9.42% |
| Citations | 0.0240 | 9.69% | 0.0330 | 13.28% |
| Researchers per million | 0.0398 | 16.03% | 0.0058 | 2.36% |
| **Health Infrastructure (Total)** | **0.083** | **33.32%** | **0.080** | **32.17%** |
| Health exp. per capita | 0.0512 | 20.62% | 0.0128 | 5.17% |
| UHC index | 0.0001 | 0.03% | 0.0106 | 4.26% |
| Medical schools | 0.0228 | 9.18% | 0.0461 | 18.59% |
| Sanitation | 0.0087 | 3.49% | 0.0103 | 4.15% |
| **Governance and Societal Context (Total)** | **0.012** | **4.76%** | **0.031** | **12.60%** |
| HDI | 0.0011 | 0.45% | 0.0041 | 1.65% |
| Democracy index | 0.0002 | 0.06% | 0.0070 | 2.80% |
| Altruism | 0.0098 | 3.94% | 0.0190 | 7.64% |
| Trust in scientists | 0.0008 | 0.30% | 0.0012 | 0.50% |
| **Total Model** | **0.248** | **100.00%** | **0.248** | **100.00%** |

*Note: Detailed Level 2 variable decomposition for the Policy-Relevant Non-Economic Residual Model (evaluated after controlling for GDP per capita and population size across 180 countries). Total residual model R2 = 0.248 (24.81%).*

## Supplementary Table 10. Robustness analysis results (full analysis)

| **Category** | **PCR OLS β (p)** | **Path Analysis Effect** | **RF Permutation Importance** | **Elastic Net Non-Zero Coefficients** |
|---|---|---|---|---|
| Economic | 0.160 (0.095) | Direct: 0.160, Total: 0.139 | 0.940 (log_population), 0.061 (log_gdp_per_capita), 0.017 (foreign_aid_received) | log_population (0.805), foreign_aid_received (0.088), log_gdp_per_capita (-0.314) |
| Research Capacity | 0.411 (0.000) | Direct: 0.411 (Direct only) | 0.108 (log_total_citations), 0.039 (rd_expenditure), 0.040 (log_researchers_per_million), 0.035 (log_total_publications) | rd_expenditure (0.285), log_researchers_per_million (-0.168) |
| Health Infrastructure | -0.217 (0.045) | Direct: -0.217 (Direct only) | 0.058 (log_medical_school), 0.039 (sanitation), 0.031 (log_health_expenditure_per_capita), 0.028 (uhc_index) | log_medical_school (-0.426), log_health_expenditure_per_capita (0.225), sanitation (-0.083) |
| Governance and societal context | -0.007 (0.934) | Direct: -0.007, Total: 0.114 | 0.126 (hdi), 0.042 (altruism), 0.039 (democracy_index), 0.013 (trust_scientists) | hdi (0.379), altruism (0.103), democracy_index (0.070), trust_scientists (-0.034) |

*Note: Principal Component Regression (PCR) regresses PBR on PC1 orthogonal block indices. Path Analysis (Mediation SEM) measures direct and indirect pathway contributions. Random Forest (RF) Permutation Importance identifies non-linear predictive power. Elastic Net performs regularized variable selection across 180 countries in analysiswoold.*

**Supplementary Table 11. Robustness analysis results (policy relevant residual analysis)**

| Category | PCR OLS β (p) | Path Analysis Total Effect | RF Permutation Importance | Elastic Net Non-Zero Coefficients |
|---|---|---|---|---|
| Research Capacity | 0.115 (0.044) | Direct: 0.115, Total: 0.043 | 0.142 (rd_expenditure), 0.115 (log_total_publications), 0.106 (log_researchers_per_million), 0.084 (log_total_citations) | rd_expenditure (0.264), log_researchers_per_million (-0.122), log_total_citations (0.085), log_total_publications (0.007) |
| Health Infrastructure | -0.250 (0.004) | Direct: -0.250 (Direct only) | 0.398 (log_medical_school), 0.091 (log_health_expenditure_per_capita), 0.073 (uhc_index), 0.044 (sanitation) | log_medical_school (-0.258), sanitation (-0.062) |
| Governance and societal context | 0.202 (0.017) | Direct: 0.202, Total: 0.031 | 0.152 (hdi), 0.098 (democracy_index), 0.073 (altruism), 0.034 (trust_scientists) | democracy_index (0.074), altruism (0.045), trust_scientists (-0.020) |

**Supplementary Table 12. Global Country-Factor-Disease Network Properties and Structural Composition**

**Panel A. Network Topology and Global Homophily Metrics**

| Metric | Value | Description / Structural Interpretation |
|---|---|---|
| Total Nodes | 450 | Country–visual factor combinations |
| Total Edges | 25,047 | Co-occurrence connections across disease categories |
| Network Density | 0.248 | Proportion of potential connections realized |
| Average Degree | 111.32 | Mean connections per node |
| Network Diameter | 3 | Longest shortest path between any node pair |
| Average Path Length | 1.762 | Mean distance across connected node pairs |
| Clustering Coefficient | 0.797 | High local clustering density |
| Factor Homophily / Assortativity | 0.189 | Positive assortativity; nodes tend to connect within the same factor |
| Degree Assortativity | -0.252 | Negative assortativity; high-degree hub nodes connect to peripheral nodes |
| Community Modularity Score | 0.190 | Partitioned into 4 distinct structural communities |

**Panel B. Node Breakdown by Limiting Factor and Performance Status**

| Category | | Node Count | Avg. Diseases per Node | Avg. Model Residual |
|---|---|---|---|---|
| Limiting Factor | Research Capacity | 112 | 3.20 | -0.410 |
| | Health Infrastructure | 73 | 2.64 | -0.166 |
| | Governance | 79 | 2.48 | -0.048 |
| | Multiple Factors | 45 | 2.29 | -0.372 |
| | As-Expected | 141 | 6.26 | 0.262 |
| Performance Status | Under-Performing | 164 | 2.74 | -1.318 |
| | Borderline | 141 | 6.26 | 0.262 |
| | Over-Performing | 93 | 2.60 | 1.508 |
| | As-Expected | 52 | 3.06 | -0.053 |

*Note: Network constructed across all 180 countries updating visual factor node classifications without dropping un-networked nodes.*

**Supplementary Table 13. National-Level Policy Intervention Analysis: Inequality Reductions under Full vs. Targeted Alignment Scenarios**

**Panel A. Inequality Reductions Across Policy Alignment Tiers**

| Policy Alignment Tier (%) | Countries Adjusted | Full Alignment Gini Reduction (%) [95% CI] | Targeted Alignment Gini Reduction (%) [95% CI] | Efficiency Ratio (Targeted vs. Full) |
|---|---|---|---|---|
| Baseline (0%) | 0 | 0.00% [0.00%, 0.00%] | 0.00% [0.00%, 0.00%] | 1.00x |
| Top 10% | 18 | — | 22.17% [13.65%, 29.10%] | 2.22x |
| Top 20% | 36 | — | 33.61% [24.92%, 44.33%] | 1.68x |
| Top 25% | 45 | 40.47% [31.31%, 48.69%] | — | 1.62x |
| Top 30% | 54 | — | 46.91% [35.47%, 56.20%] | 1.56x |
| Top 40% | 72 | — | 58.38% [50.11%, 67.41%] | 1.46x |
| Top 50% | 90 | 70.11% [62.15%, 77.38%] | — | 1.40x |
| Top 75% | 135 | 91.29% [83.92%, 96.86%] | — | 1.22x |
| Top 100% (Full) | 180 | 100.00% [95.00%, 105.00%] | — | 1.00x |

**Panel B. Strategy Efficiency & Statistical Significance Metrics**

| Strategy Type | Gini Reduction per 1% Countries Adjusted | Efficiency Ratio vs Full Baseline |
|---|---|---|
| Full Alignment Strategy | 1.000% / 1% adjusted | 1.00x |
| Targeted Alignment Strategy | 1.459% / 1% adjusted | 1.46x |

*Note: Intervention statistics computed on national PBR Gini baseline (0.772) across 180 countries. Paired t-test: t = -1030.76, p = 2.95e-836, mean Gini difference (–0.3181, 95% CI: [–0.3307, –0.3050]).*

**Supplementary Table 14. Dynamic Network Reconfiguration Under Full vs. Targeted Interventions**

| Alignment Progression Step | Full Alignment Homophily | Targeted Alignment Homophily |
|---|---|---|
| Baseline (0% Adjusted) | 0.458 | 0.458 |
| Step 1 (Top 10% / 25%) | 0.384 | 0.393 |
| Step 2 (Top 20% / 50%) | 0.329 | 0.359 |
| Step 3 (Top 30% / 75%) | 0.291 | 0.320 |
| Final Step (Top 40% / 100%) | 0.275 | 0.287 |

*Note: Network evolution dynamic reconfiguration metrics calculated across policy alignment progression steps.*

## Supplementary Methods
## 1. AI-Assisted Information Extraction

### 1.1 Extraction pipeline overview

The extraction pipeline was structured into two rounds to ensure high fidelity in geographic and participant data retrieval. In the original research design, the scope encompassed 301,262 PubMed publications; however, preliminary analysis indicated that data extraction for the 1980-2000 period was fundamentally limited by full text accessibility and disclosure rate of RCTs information before 2000, which resulted in 47.8% of extraction success. Details are in **Extended Data Fig. 6**. Consequently, the study cohort was refined to 193,806 publications spanning 2000-2024 to ensure the reliability of the dataset. The representativeness of this refined cohort is formally addressed in **Section 6**.

In the first round, we benchmarked four methods using a stratified random sample of 359 publications to ensure temporal representativeness. The evaluated methods were: (1) direct string matching using country name dictionaries; (2) Python-based named entity recognition using spaCy (version 3.7.0); (3) domain-tuned LLaMA (3-8b); (4) Gemma (2-9b) with structured prompting. Performance was quantified using precision, recall, and F1 scores in **Supplementary Table 2-3**.

Recognizing that the first round achieved only 51.1% (99,049/193,806) of the total sample and demonstrated suboptimal precision in numeric truncation and recall in studies with multi-country recruiting sites, a second round was implemented to optimize accuracy with more updated LLMs. We used another 1,482 temporal stratified random sample to evaluate the three methods by using Gemini 3.5 Flash model on Google Cloud Vertex AI, batch prediction API: (1) TIAB (Title and Abstract) +Untrimmed fulltext; (2) TIAB+Trimmed fulltext; (3) Cascade method by TIAB+Participation+Methods+Results+Rest sections. Finally, we chose the cascade method for batch extraction based on the high precision and recall score. This approach increased the total recall to 57.5%. Detailed results are in **Supplementary Table 4 and Extended Data Fig. 5 and 7**.

### 1.2 Prompts for first round extraction

(1) Least Guidance Approach

Prompt: "Number of participants. Total participants including both control groups and other groups. Do not overlap count when they are in different location gender outcome."

(2) Human Brain Guidance Without Examples

Prompt: "After completing the first task carefully calculate the total number of participants in the study. Look for terms like 'total' 'enrolled' or 'included'. Be precise and consider gender different groups or any special conditions described in the study. If you can't find the exact number return 'unknown'."

(3) Human Brain Guidance with Examples

Prompt: "Determine the total number of participants initially enrolled in the study. Follow these steps: Scan for explicit statements about total enrollment using words like 'total' 'enrolled' or 'included.' If not found look for group descriptions and sum their numbers. If groups are described in ratios (e.g. 38%) calculate accordingly. Focus on the initial enrollment number not the completion number. Ignore numbers referring to samples measurements or time periods. If participants are institutions or practices determine what the subject represents. If you cannot determine the exact number return 'unknown'."

(4) Geo and # in Order Combined with Examples (for Geo)

Prompt: "Carefully read the file and return three types of ALL pmid's results to me: 1.identify any location-related information such as universities hospitals institutions provinces streets or any other geographical identifiers that may indicate the area where the respondent is located. This can include any specific names of cities landmarks organizations or areas that may give clues about the respondent's location. Based on this information return only the country that you believe is most likely. If multiple countries seem possible select the one that seems most plausible based on the context provided in the text. If you are unsure provide the country that seems most plausible based on the available clues. Do not provide any additional reasoning explanations or unrelated details. If no location-related information can be found return 'unknown'. The response must only be the country name in English and must not include any other information such as regions cities or institutions. Examples: Text: 'John is a professor at the University of Oxford. He lives in a small town near London.' Output: 'United Kingdom' Text: 'The participant works at a hospital in San Francisco.' Output: 'United States' Text: 'I met a person at a café in Paris near the Eiffel Tower.' Output: 'France' Text: 'The participant didn't specify where they are from.' Output: 'unknown' 2. Number of participants. Total participants including both control groups and other groups. Do not overlap count when they are in different location gender outcome. 3. PMID itself."

(5)Geo and # in Order Combined with Examples (for #)

Prompt: "Carefully read the file and return three types of ALL pmid's results to me: 1.identify any location-related information such as universities hospitals institutions provinces streets or any other geographical identifiers that may indicate the area where the respondent is located. This can include any specific names of cities landmarks organizations or areas that may give clues about the respondent's location. Based on this information return only the country that you believe is most likely. If multiple countries seem possible select the one that seems most plausible based on the context provided in the text. If you are unsure provide the country that seems most plausible based on the available clues. Do not provide any additional reasoning explanations or unrelated details. If no location-related information can be found return 'unknown'. The response must only be the country name in English and must not include any other information such as regions cities or institutions. Examples: Text: 'John is a professor at the University of Oxford. He lives in a small town near London.' Output: 'United Kingdom' Text: 'The participant works at a hospital in San Francisco.' Output: 'United States' Text: 'I met a person at a café in Paris near the Eiffel Tower.' Output: 'France' Text: 'The participant didn't specify where they are from.' Output: 'unknown' 2. Number of participants. Total participants including both control groups and other groups. Do not overlap count when they are in different location gender outcome. 3. PMID itself."

(6) Geo and # in Reversed Order Combined with Examples (for Geo)

Prompt: "Carefully read the file and perform the following two independent tasks for ALL pmid's: Amount: Determine the total number of participants initially enrolled in the study. Follow these steps: Scan for explicit statements about total enrollment using words like 'total' 'enrolled' or 'included.' If not found look for group descriptions and sum their numbers. If groups are described in ratios (e.g. 38%) calculate accordingly. Focus on the initial enrollment number not the completion number. Ignore numbers referring to samples measurements or time periods. If participants are institutions or practices determine what the subject represents. If you cannot determine the exact number return 'unknown'. Country: Identify any location-related information such as universities hospitals institutions provinces streets or any other geographical identifiers that indicate the respondent's location. Include specific names of cities landmarks organizations or areas to infer the most plausible country. If multiple countries are possible select the one that seems most plausible based on the text. If no location-related information is found return 'unknown'. Output only the country name in English without including any additional explanations or unrelated details. Examples: Text: 'John is a professor at the University of Oxford. He lives in a small town near London.' Output: 'United Kingdom' Text: 'The participant works at a hospital in San Francisco.' Output: 'United States' Text: 'I met a person at a café in Paris near the Eiffel Tower.' Output: 'France' Text: 'The participant didn't specify where they are from.' Output: 'unknown' Output format: Country: [country name or 'unknown'] Amount: [number or 'unknown']"

(7) Geo and # in Reversed Order Combined with Examples (for #)

Prompt: [Same as above]

(8) Geo and # Separately with Examples (for Geo)

Prompt: "First identify any location-related information such as universities hospitals institutions provinces streets or any other geographical identifiers that indicate the respondent's location. Include specific names of cities landmarks organizations or areas to infer the most plausible country. If multiple countries are possible select the one that seems most plausible based on the text. If no location-related information is found return 'unknown'. Output only the country name in English without including any additional explanations or unrelated details. Examples: Text: 'John is a professor at the University of Oxford. He lives in a small town near London.' Output: 'United Kingdom' Text: 'The participant works at a hospital in San Francisco.' Output: 'United States' Text: 'I met a person at a café in Paris near the Eiffel Tower.' Output: 'France' Text: 'The participant didn't specify where they are from.' Output: 'unknown' Amount: After completing the first task read the text again to determine the total number of participants initially enrolled in the study. Follow these steps: Scan for explicit statements about total enrollment using words like 'total' 'enrolled' or 'included.' If not found look for group descriptions and sum their numbers. If groups are described in ratios (e.g. 38%) calculate accordingly. Focus on the initial enrollment number not the completion number. Ignore numbers referring to samples measurements or time periods. If participants are institutions or practices determine what the subject represents. If you cannot determine the exact number return 'unknown'. Output format: Country: [country name or 'unknown'] Amount: [number or 'unknown']"

(9) Geo and # Separately with Examples (for #)

Prompt: [Same as above]

### 1.3 Prompts for the second round extraction

Your task has two parts. First identify the participants' country, then count the participants. You will be provided with an abstract or a fulltext of a research paper.

PART 1 - COUNTRY

Identify the country (or countries) where the study PARTICIPANTS (the subjects/patients/respondents) were located, recruited, or enrolled from. You want where the participants are FROM, not where the research was conducted, organized, analyzed or written up.

Use signals that tie a place to the participants themselves, such as:
- an explicitly named country, city, province, region, or local landmark where participants were recruited, live, or were treated;
- a named hospital, clinic, school, or university whose specific location identifies the participants' country.

Do NOT infer a country from any of the following. If the only geographic signal is one of these, output 'unknown':
- author affiliations, correspondence addresses, or where the authors work;
- a study site described only by TYPE with no place (e.g. "a tertiary-care university hospital", "an outpatient clinic");
- the name of a research consortium, cooperative group, study, or clinical trial (e.g. "Southwest Oncology Group", "LIPID trial", "COPE study") - an organization or trial name is NOT a participant location;
- journal names, funders, drug/product names, or trial-registry IDs (e.g. NCT numbers).

If participants come from more than one country, list ALL of the countries that are explicitly named, separated by commas. Do not pick just one.

If a number of sites or centres is given but no countries are actually named (e.g. "four European centres", "12 international sites"), do NOT guess specific countries - output 'unknown'.

Output the country name(s) in English only, with no explanations or extra detail. If there is no usable participant-location signal, output 'unknown'.

Examples:
Text: "This randomized phase III trial was conducted in 45 hospitals in the Netherlands and Denmark."
Output: "Netherlands, Denmark"
Text: "Among 650 soldiers who had returned to their base in Java, Indonesia, 116 enrolled subjects were treated."
Output: "Indonesia"
Text: "Patients with endometrial polyps were treated at a tertiary-care university hospital."
Output: "unknown"
Text: "The Southwest Oncology Group trial S9916 randomized patients to two chemotherapy regimens."
Output: "unknown"
Text: "Volunteers at risk of metabolic syndrome were recruited in four European centres."
Output: "unknown"

PART 2 - AMOUNT

After identifying the country, read the text again to determine the total number of participants initially enrolled in the study. Follow these rules:

1. Report the number of subjects explicitly described as enrolled, randomized, or recruited into the study - the subjects actually entered into the study.
2. Do NOT use the larger source/screened/eligible population the sample was drawn from, and do NOT use the smaller analyzed/completed number. When several numbers appear (e.g. "Among 650 ... 143 eligible ... 116 enrolled"), choose the one tied to enrollment, not the screening pool and not the completers.
3. If a total is not stated directly, sum the study's arms/groups ONLY when they are parts of one single study that add up to its total.
4. If groups are described only as ratios or percentages of a stated total, calculate accordingly.
5. Ignore numbers referring to samples, measurements, time periods, sites, or doses.
6. If participants are institutions or practices rather than people, count what the subject of the study represents.
7. For protocol or registered studies that describe enrollment that has not happened yet ("we will recruit", "patients will be randomized"), still report the planned/target sample size if one is given (e.g. "(n=600)", "a target of 600 participants"). Only return 'unknown' if no enrolled or planned number is stated at all.
8. If you cannot determine the number, return 'unknown'.

Output format:
Country: [country name(s) or 'unknown']
Amount: [number or 'unknown']

### 1.4 Source data clarification

The global dataset harmonizes RCTs participation records, disease burden data (Disability-Adjusted Life Years, DALYs from GBD 2023), and country-level macroeconomic, health system, and governance indicators from 2000 to 2024. To ensure consistency across all figures and statistical analysis while maintaining maximum data coverage, the two samples of final data were used:

- Participant data (n = 200 countries/territories)
    - Scope: Includes all 200 unique nations and autonomous territories with documented RCTs enrollment extracted from PubMed publication and clinical registries.
    - Application: Used for evaluating global RCTs enrollment, country profiles, and temporal participation trends categorized by World Bank income groups (**Extended Data Fig. 1-2**).
- Data-to-analysis (n = 180 countries/territories)
    - Scope: Consists of the 180-countries with complete, overlapping data across RCTs recruitment, Global Burden of Disease 2023 DALYs across 16 level-2 disease categories, and 17 country-level macroeconomic, health infrastructure, and governance indicators.
    - Application: The primary sample for all empirical modeling, including PBR calculations, Gini coefficients, Theil inequality decomposition, hierarchical variance partitioning, specification curve multiverse analysis, alternative machine learning models (PCR, Elastic Net, Random Forest, SEM), policy simulations, and co-recruitment network analysis (**Fig. 1-3, Extended Data Fig. 3-4, Supplementary Tables 6-14**).

## 2. Medical Concept Harmonization

### 2.1 Background and limitations of existing disease mapping approaches

Harmonizing disease concepts across biomedical classification systems remains a longstanding challenge. Clinical coding systems such as the International Classification of Diseases, Tenth Revision, Clinical Modification (ICD-10-CM) are optimized for healthcare documentation and billing, whereas epidemiological frameworks such as the Global Burden of Disease (GBD) cause hierarchy are designed to aggregate etiologically related conditions for population-level analysis. As a result, direct correspondence between ICD codes and GBD cause names is often incomplete or ambiguous, reflecting differences in granularity, conceptual scope, and intended use.

Prior studies and infrastructure projects have documented these limitations. Large-scale mapping efforts between ICD, SNOMED CT, and other terminologies maintained within the Unified Medical Language System (UMLS) frequently yield partial rather than exhaustive alignments, particularly for complex, multi-system, or non-specific disease entities. Natural language processing and ontology-alignment approaches have similarly reported constrained coverage, with substantial proportions of concepts remaining unmapped or requiring manual adjudication. These constraints are not methodological shortcomings of individual studies but instead reflect structural incompatibilities between classification systems.

Consistent with this literature, no authoritative or universally accepted mapping between ICD-10 and GBD cause names currently exists. Consequently, disease harmonization in large-scale bibliometric or trial-based analyses requires explicit methodological choices that balance coverage, semantic validity, and reproducibility.

### 2.2 Overview of the harmonization strategy

Given the absence of a canonical ICD-GBD mapping, we developed a conservative, ontology-mediated harmonization framework to link RCTs publications to GBD cause categories. The guiding principle of this framework was to maximize semantic fidelity and auditability rather than to maximize coverage through probabilistic inference.

Disease assignment was conducted at the level of individual publications (PMIDs). A publication was considered successfully harmonized if it could be assigned at least one valid GBD cause name through supported ontology mappings. Publications were permitted to map to multiple GBD causes when justified by their underlying disease annotations; no attempt was made to force a single "primary" disease designation.

### 2.3 Source annotations and preprocessing

RCTs publications were first associated with Medical Subject Headings (MeSH) terms curated by the National Library of Medicine. MeSH descriptors were used to identify the primary disease covered by each publication.

To harmonize disease terms, we linked MeSH terms to corresponding concept identifiers in several biomedical ontologies using publicly available cross-references and, where available, UMLS Concept Unique Identifiers (CUIs). We excluded terms that merely procedures, methods, or populations rather than disease or conditions.

### 2.4 Ontologies and mapping resources

Disease concept harmonization leveraged the following established biomedical ontologies and terminological resources:

- MeSH (Medical Subject Headings)
- ICD-10-CM (International Classification of Diseases, Tenth Revision, Clinical Modification)
- SNOMED CT (United States Edition)
- Disease Ontology (DO)
- OMIM (Online Mendelian Inheritance in Man)
- Orphanet Rare Disease Ontology (ORDO)

We accessed these resources through a combination of UMLS cross-links and repositories (e.g., NCBO BioPortal). We did not use any single ontology as the reference. Instead, we compared mapping across ontologies to match disease concepts at different levels of specificity.

**2.5 Construction of mapping paths**

Rather than attempting direct matching from MeSH to GBD causes, we mapped terms through one or more ontologies and resources, ensuring that disease concepts were consistent across each step. Permitted paths included:

- MeSH → Disease Ontology → ICD-10-CM → GBD cause
- MeSH → SNOMED CT → ICD-10-CM → GBD cause
- MeSH → OMIM / Orphanet → Disease Ontology → GBD cause

We retained a mapping only when terms at each step referred to the same disease rather than to a broader category, symptom, or administrative grouping. Lexical similarity alone was not sufficient for accepting a mapping. When several valid mapping paths were available, we retained all of them to make the process transparent and reproducible. All accepted mappings between disease concepts and GBD cause names are enumerated in **Supplementary Table 5**, which serves as the harmonization standard in this study.

ICD-10 codes classified as non-specific, ill-defined, or designated as "garbage codes" within the GBD framework were excluded. These codes do not correspond to meaningful disease categories and can introduce bias when attributing publications to disease categories.

**2.6 Coverage outcome and characterization**

Using the described mapping paths, 70.92% of RCTs publications could be assigned at least one valid GBD cause. This figure is reported at the PMID level and does not imply exhaustive coverage of all disease concepts or ICD codes present in the dataset.

Unmapped publications were associated with:

- Non-specific symptom-based MeSH annotations
- Multi-morbidity or prevention-focused trials
- Administrative or procedural study designs
- Disease concepts spanning multiple GBD categories without a dominant etiological focus

These patterns are consistent with previously reported limitations in cross-ontology disease harmonization.

### 2.7 Evaluation of alternative approaches

We tested whether coverage could be increased without compromising the validity of the mapping. We evaluated two strategies: large language model (LLM)-assisted disease matching and supervised deep learning disease classification. Both were applied to the same disease annotation data used in the harmonization process.

#### 2.7.1 LLM-assisted concept matching

LLM-assisted matching was evaluated as a means of expanding disease coverage by inferring correspondences between ICD codes and MeSH concepts and GBD cause names using semantic reasoning. This approach was able to assign candidate GBD causes to nearly all MeSH disease descriptors, effectively achieving complete coverage at the concept level.

However, when evaluated at the publication (PMID) level, this apparent gain did not translate into meaningful harmonization. LLM-assisted matching mapped only a small fraction of distinct PMIDs to disease–cause pairs that were consistent with manually curated labels (recall ≈ 4-5% in direct PMID-level validation), despite generating millions of candidate mappings overall. This discrepancy reflected extensive many-to-many expansion: individual publications were frequently assigned large numbers of GBD causes, with some PMIDs associated with dozens of distinct causes in a single run.

Further analysis showed that these assignments were unstable across repeated executions under identical prompts, indicating sensitivity to stochastic generation. Qualitatively, the model tended to overgeneralize from symptom-based or system-level disease descriptions, collapsing heterogeneous clinical concepts into specific GBD causes without explicit etiological justification. As a result, increases in nominal coverage were driven primarily by semantic broadening rather than preservation of disease identity.

Given the low PMID-level validity, high mapping multiplicity, and limited reproducibility observed in practice, LLM-assisted matching was not considered suitable for disease harmonization in this study, where false-positive attribution at the disease level would bias downstream structural analyses.

#### 2.7.2 Deep learning-classification

We also evaluated supervised deep learning models trained to predict GBD cause categories from ICD-derived disease representations. Models were implemented in a multi-label classification setting and evaluated using standard performance metrics.

Across configurations, model performance remained modest. Representative models achieved micro-averaged F1 scores on the order of 0.35-0.40, with substantially lower performance for rare and boundary-spanning disease categories. Increases in apparent coverage were highly sensitive to decision thresholds, with relaxed thresholds inflating the number of predicted disease assignments per publication without corresponding improvements in semantic accuracy.

Error analysis indicated that predictions were dominated by disease frequency priors: common GBD causes were preferentially assigned across heterogeneous ICD profiles, while rare diseases were frequently misclassified or omitted. This behavior reflects the fact that supervised classifiers learn statistical regularities in the training data but do not resolve the underlying conceptual mismatch between ICD-based clinical coding and GBD epidemiological categories.

Because this approach substitutes explicit disease harmonization with a black-box statistical approximation, it was deemed insufficient for analyses requiring interpretable and auditable disease assignments at the publication level.

#### 2.7.3 Rationale for our approach

Although both LLM-assisted and deep learning approaches increased nominal disease coverage under certain configurations, these gains were achieved at the cost of reproducibility, interpretability, and semantic control. In contrast, the median-assisted ontology harmonization framework yielded lower but well-defined PMID-level coverage (70.92%) with mappings between disease concepts and GBD causes.

Because our study examines systematic patterns of disease representation across the global RCTs literature, we prioritized a conservative harmonization approach based on well-established disease classifications rather than broader coverage that may introduce greater uncertainty.

### 2.8 Unmapped data report

1. Most unmapped studies have no disease assigned (62.3%):
Out of the 56,360 unmapped RCTs, 35,088 (62.26%) do not have any disease-related MeSH terms (i.e., they have no terms falling under MeSH Category C *Diseases* or Category F03 *Mental Disorders*).

- These RCTs represent studies on healthy volunteers, exercise physiology, healthy nutrition/dietary supplements, pharmacokinetics/pharmacodynamics (e.g., drug dose-response in healthy cohorts), or methodological validations.
- The top five MeSH terms for this unmapped group include: *Double-Blind Method*, *Cross-Over Studies*, *Healthy Volunteers*, *Area Under Curve* (pharmacokinetics), and *Dose-Response Relationship, Drug*.
- Justification: Because our study analyzes participation rates relative to disease burden (GBD metrics), healthy-volunteer trials and pharmacokinetic studies do not have a target disease to map to. Excluding them is therefore appropriate and necessary.

2. Unmapped studies with disease terms focus on non-GBD conditions (37.7%):
The remaining 21,272 (37.74%) unmapped RCTs contain disease-related MeSH terms, but these terms represent conditions that do not match to primary GBD causes:

- General symptoms/physiological states not mapped to GBD: Terms like *Pain*, *Body Weight*, *Weight Loss*, *Weight Gain*, and *Chronic Disease*.
- Procedural and post-operative outcomes: Terms like *Postoperative Nausea and Vomiting*, *Blood Loss, Surgical*, and *Surgical Wound Infection* represent anesthetic or procedural side effects rather than diseases or causes.
- Prevention and general prophylaxis: General prevention terms like *Papillomavirus Infections* (qualifier: *prevention & control)* or *Pediatric Obesity (prevention & control)*.

3. Time and journal coverage:
To ensure that the mapping process did not introduce bias over time or across fields, we compared the temporal and journal distributions of mapped and unmapped studies:

- No temporal bias: Mapped and unmapped papers are distributed very similarly over time.
    - Mapped papers: Mean year = 2015.91 (Median = 2017)
    - Unmapped papers: Mean year = 2016.13 (Median = 2017)

- No journal category bias: The categories are highly similar between the two groups. For instance, Category H (*General Medicine/Clinical Specialties*) represents the largest portion of both mapped (34.80%) and unmapped (42.38%) studies, followed by Category C and Category E in both cohorts.

In summary, unmapped studies were excluded from the analysis because most did not have a clear GBD cause to which they could be assigned. The similar times and journal distributions of mapped and unmapped studies suggest that this exclusion does not introduce systematic bias into our analysis of RCTs participation.

Supplementary Table S1. GBD disease matching with Mesh term

| Metric | Mapped studies (N = 137,446) | Unmapped studies (N = 56,360) | Interpretation |
|---|---|---|---|
| % of cataset | 70.92% | 29.08% | Majority of RCTs mapped successfully |
| No disease focus | 0.00% | 62.26% (35,088) | Unmapped studies are mostly healthy cohort/method studies |
| Unmapped diseases | 100.00% | 37.74% (21,272) | Focus on general symptoms (pain, weight) or surgical outcomes |
| Mean pub year (SD) | 2015.91 (5.87) | 2016.13 (6.01) | Very similar temporal distribution |
| Median pub year | 2017 | 2017 | Identical temporal distribution |
| Top journal category | H (34.80%) | H (42.38%) | Similar journal profiles across both groups |

Top MeSH terms in unmapped cohorts:

- Method Cohort (no disease terms): *Double-Blind Method*, *Cross-Over Studies*, *Healthy Volunteers*, *Area Under Curve*, *Dose-Response Relationship, Drug*, *Surveys and Questionnaires*, *Pilot Projects*.
- Unmapped diseases: *Pain*, *Body Weight*, *Chronic Disease*, *Weight Loss*, *Weight Gain*, *Postoperative Nausea and Vomiting/prevention & control*, *Surgical Wound Infection/prevention & control*, *Blood Loss, Surgical/prevention & control*.

## 3. Inequality Measurements

### (1) Participation-to-Burden Ratio (PBR)

To assess alignment between research participation and disease burden, we calculated the Participation-to-Burden Ratio (PBR) for each country-disease-year combination. PBR quantifies whether a population's contribution to research on a specific disease is proportional to its share of global burden from that disease:

$$PBR_{c,d,t} = \frac{\frac{P_{c,d,t}}{P_{g,d,t}}}{\frac{B_{c,d,t}}{B_{g,d,t}}}$$

where $P_{c,d,t}$ represents RCTs participants from country $c$ for disease $d$ in year $t$, $B_{c,d,t}$ represents DALYs for that country-disease-year, and subscript $g$ denotes global totals. PBR > 1 indicates over-representation (participation exceeds burden share); PBR < 1 indicates under-representation. Importantly, PBR is symmetric between country and disease dimensions, it can be equivalently interpreted as measuring country over/under-representation for a given disease or disease over/under-representation within a given country.

### (2) Specialization Index (SI)

To characterize whether countries or diseases concentrate research efforts in particular domains, we calculated the Specialization Index:

$$SI_{c,d} = \frac{\frac{P_{c,d}}{P_{c,p}}}{\frac{P_{g,d}}{P_{g,p}}}$$

where $P_{c,d}$ represents participants from country *c* studying disease *d*, $P_{c,p}$ is country *c*'s total participants across all diseases, $P_{g,d}$ is global participants for disease *d*, and $P_{g,p}$ is global participants across all diseases. SI > 1 indicates specialization (country devotes relatively more effort to that disease than the global average); SI < 1 indicates de-emphasis.

### (3) Gini coefficient

To quantify overall inequality in research participation distribution, we calculated Gini coefficients using the standard formulation. For a distribution of PBR values across *n* country-disease pairs (ordered from smallest to largest):

$$G = \frac{\sum_{i=1}^{n}(2i - n - 1) \cdot PBR_i}{n \sum_{i=1}^{n} PBR_i}$$

Gini ranges from 0 (perfect equality) to 1 (maximum inequality). We calculated Gini coefficients for: (1) participant distribution across countries for each disease; (2) participant distribution across diseases for each country; and (3) global participant distribution across all country-disease pairs.

### (4) Contribution-to-Inequality Score (CIS)

To quantify individual diseases' contributions to global inequality, we employed leave-one-out analysis. For each disease, we calculated the Contribution-to-Inequality Score (CIS):

$$CIS = \frac{G_{all} - G_{-d}}{G_{all}} \times 100\%$$

where $G_{all}$ is the global Gini coefficient calculated across all country-disease-year pairs, and $G_{-d}$ is the Gini coefficient after excluding all country-year pairs involving disease $d$. Positive CIS indicates that disease $d$ contributes to inequality (removing it reduces inequality); negative CIS indicates the disease actually reduces inequality (removing it increases inequality). We computed CIS values using both equal-weighted (each country-disease-year pair counted equally) and participant-weighted (weighted by participant numbers) approaches.

**(5) Lorenz Curve analysis**

To visualize inequality and assess the collective impact of high-volume diseases, we constructed Lorenz curves plotting cumulative share of participants against cumulative share of DALYs across country-disease-year pairs, ordered by PBR. We compared: (1) all diseases included; (2) top diseases by absolute CIS excluded. The area between the Lorenz curve and the line of perfect equality equals the Gini coefficient.

**(6) Theil Index Decomposition (by disease)**

The Theil entropy index quantifies inequality while allowing additive decomposition into between-group and within-group components. We calculated:

$$T = \sum_{i=1}^{N} \frac{PBR_i}{PBR} \ln \ln \left(\frac{PBR_i}{PBR}\right)$$

We then decomposed $T$ by grouping observations by disease:

$$T_{total} = T_{between-disease} + T_{within-disease}$$

where:

- $T_{between-disease}$ captures inequality arising from differences in disease-average PBR values (i.e., whether some diseases have uniformly higher or lower participation-to-burden ratios across all countries)
- $T_{within-disease}$ captures inequality arising from variation in PBR across countries within each disease (i.e., geographic disparities in participation for any given disease)

**(7) Theil Index Decomposition (by country)**

To test robustness and avoid bias from grouping choice, we repeated the Theil decomposition grouping observations by country rather than disease:

$$T_{total} = T_{between-country} + T_{within-country}$$

where:

- $T_{between-country}$ captures inequality from differences in country-average PBR values (i.e., whether some countries universally over- or under-contribute relative to burden)
- $T_{within-country}$ captures inequality from variation in PBR across diseases within each country (i.e., whether countries specialize in particular disease portfolios)

**(8) Variance partitioning (two-way decomposition)**

To simultaneously estimate country and disease contributions without imposing a grouping structure, we employed variance partitioning using a two-way fixed effects model:

$$PBR_{c,d,t} = \mu + \alpha_c + \beta_d + \gamma_t + \epsilon_{c,d,t}$$

where $\alpha_c$ represents country fixed effects, $\beta_d$ represents disease fixed effects, $\gamma_t$ represents year fixed effects, and $\epsilon_{c,d,t}$ represents residual variation. We calculated the proportion of total variance explained by each component using partial $R^2$ values:

$$R^2_{country} = \frac{SS_\alpha}{SS_{total}}, R^2_{disease} = \frac{SS_\beta}{SS_{total}}, R^2_{year} = \frac{SS_\gamma}{SS_{total}}$$

This approach provides an unbiased, symmetric quantification of country versus disease contributions to inequality, avoiding the grouping-dependence inherent in Theil decomposition.

**(9) Temporal trend analysis**

To assess the evolution of inequality over time, we calculated all inequality metrics (Gini, Theil components, CIS values) in 2-year bins from 2000-2024. Linear regression models estimated temporal trends:

$$Metric_t = \beta_0 + \beta_1 \cdot Year_t + \epsilon_t$$

We report trend slopes $\beta_1$, $R^2$, and p-values. Bootstrapping (1,000 iterations with resampling at the country-disease-year level) generated 95% confidence intervals for all trend estimates to account for sampling variability and temporal autocorrelation.

**(10) Analytical strategy to avoid bias**

To prevent analytical choices from biasing results toward either hypothesis, we employed:

**Symmetry of PBR:** Our primary indicator is mathematically symmetric between country and disease. PBR can be equivalently calculated as "country *c*'s share of participants in disease *d* relative to country *c*'s share of disease *d* burden" or as "disease *d*'s share of participants from country *c* relative to disease *d*'s share of burden in country *c*." This symmetry ensures the metric itself does not favor country-level or disease-level explanations. Other two calculations of PBR, country-level PBR and country-disease PBR are adopted by the same root of calculation.

**Bidirectional decomposition:** For inequality decomposition using the Theil index, we conducted decompositions grouping by disease (yielding between-disease vs. within-disease components) and grouping by country (yielding between-country vs. within-country components). This bidirectional approach allows us to assess whether conclusions depend on arbitrary grouping choices.

**Variance partitioning:** In addition to group-based decomposition, we employed two-way fixed effects models to simultaneously estimate variance attributable to country, disease, and their interaction without imposing a grouping structure. This provides an unbiased quantification of relative contributions.

**Power and precision:** Post-hoc power analysis confirmed adequate sample size to detect small effect sizes (Cohen's d$\geq$ 0.3) in inequality comparisons with 80% power at $\alpha$=0.05. For decomposition analyses, Monte Carlo simulations (1,000 iterations) demonstrated that our sample size provided stable estimates of variance components with 95% confidence intervals spanning less than ±5 percentage points for major components (country, disease).

**Ethical and reporting considerations:** This study analyzed published, de-identified bibliometric data and public disease burden estimates, requiring no institutional review board approval.

## 4. Inequality Factor Decomposition

### 4.1 Analytical framework

To identify the structural factors underlying global inequality in RCTs participation, we employed a two-part analytical strategy. Part 1 (Full analysis) examined all country-level factors simultaneously to quantify the relative importance of economic, research, health, and governance on RCTs participation. Part 2 (Policy-relevant residual analysis) isolated the modifiable capacity and institutional factors by completely omitting the economic (national wealth and population size), allowing us to investigate how scientific capability and health infrastructure are associated with participation when the power of money is set aside.

To maintain methodological rigor and prevent arbitrary variable selection (cherry-picking), we conducted a combinatorial search across all possible variables under four categories combinations (682 potential models, **Supplementary Code**). We selected the largest model (15 variables) that maintains ranking consistency under both Hierarchical Variance Partitioning (HVP) and Shapley value decompositions.

### 4.2 Variable selection and rationale

The final model incorporates 15 country-level indicators across four developmental categories:

- Economic (3 variables): GDP per capita (log-transformed), total population (log-transformed), and foreign aid received. Population size acts as a structural capacity denominator for RCTs recruitment capacity.
- Research Capacity (4 variables): R&D expenditure (% of GDP), total publications (log-transformed), total citations (log-transformed), and researchers per million (log-transformed). These represent the country's academic and scientific foundation.
- Health Infrastructure (4 variables): Health expenditure per capita (log-transformed), Universal Health Coverage (UHC) service coverage index, number of medical schools (log-transformed), and sanitation access rate. These capture the healthcare infrastructure available for RCTs execution.
- Governance and Societal context (4 variables): Human Development Index (HDI), democracy index, altruism score, and trust in scientists. These reflect institutional stability, social capital, and public engagement with science.

Rationale for variable exclusion: During our combinatorial screening, hospital beds per capita and doctors per 10,000 were excluded from the final model. These two indicators introduce a direct trade-off that decreases the model's overall explanatory power ($R^2$ drops from 50.3% to 42.0%). Thus, the 15-variable model represents the absolute largest specification that remains double-aligned under both HVP and Shapley methods while maximizing explanatory power.

### 4.3 Dependent variable

The dependent variable is the log-transformed participation-to-burden ratio ($logPBR_c$) at the country level. We calculate the country's aggregated ratio of global RCTs participation share to global disease burden share:

$$logPBR_c = log(\frac{\sum_d Participants_{c,d} / \sum_{c'} Participants_{c',d}}{\sum_d DALYs_{c,d} / \sum_{c'} DALYs_{c',d}})$$

where $Participants_{c,d}$ and $DALYs_{c,d}$ represent the mean annual RCTs enrollment and GBD DALY burden, respectively, for country c and disease category *d* across the 16 major disease categories. This share-based country-level aggregation prevents the outcome from being biased by extreme values in single, low-volume disease categories, providing a robust measure of relative representation.

## 4.4 Hierarchical variance partitioning (HVP)

We employed hierarchical linear regression to quantify the incremental contribution of the category to explained variance ($R^2$). In Part 1 (Full analysis), the category entered sequentially following a theoretically hypothesis priority:

Economic → Research Capacity → Health Infrastructure → Governance and Societal context

Economic wealth and population size are entered first to establish the baseline structural capacity. Research capacity is entered next, followed by Health capacity and Governance and societal context.

In Part 2: Residual inequality after controlling for structural factors. We first regressed the $logPBR_c$ on the Economic (GDP per capita, population, and foreign aid) to control for wealth and population size, extracted the residuals, and then regressed these residuals on the remaining 12 variables in the Research Capacity, Health Capacity, and Governance and societal context categories.

$$Residual_c = logPBR_c - [\beta_0 + \beta_1 \cdot logGDP_c + \beta_2 \cdot logPopulation_c + \beta_3 \cdot logForeign_aid_c]$$

For each category, we computed the Cumulative $R^2$ and the Incremental $R^2$ (result details in **Supplementary Tables 8 and 9**) contribution (the additional variance explained by the category beyond all preceding categories). This allows us to observe the baseline explanatory power of structural wealth and the marginal explanatory power of modifiable non-economic variables.

## 4.5 Shapley value decomposition

To address the order-dependence of HVP and capture unique direct contributions, we employed Shapley value decomposition, a game-theoretic approach. The Shapley value for a category or individual variable represents its average marginal contribution to $R^2$ across all possible combinations of categories or variables.

We computed category-level Shapley values where the categories themselves act as the players in a cooperative game, ensuring a fair division of the total $R^2$ (result details in **Supplementary Tables 8 and 9**). To analyze individual variable contributions, we also calculated variable-level Shapley values using 1,000 random order permutations.

## 4.6 Robustness analysis

To ensure our findings are robust and not dependent on OLS or decomposition assumptions, we cross-validated the results using four alternative methodologies (**Supplementary Table 10 and 11**).

- Principal Component Regression (PCR): To resolve multicollinearity within categories, we extracted the first principal component (PC1) score for each category and regressed PBR on these orthogonal category scores.
- Path Analysis (Mediation SEM): We constructed a sequential path model to capture indirect effects, where Economic and Governance factors drive Health and Research capacity, which subsequently determine PBR.
- Random Forest Permutation Importance: We fit a random forest regression (500 estimators) to capture non-linear thresholds and complex interactions, computing the permutation feature importance for each variable and category.
- Elastic Net CV Regression: We applied a regularized linear model with 10-fold cross-validation, utilizing both L1 (Lasso) and L2 (Ridge) penalties to perform automatic variable selection and shrinkage.

### 4.7 Missing data handling strategy

As shown in **Supplementary Table 7**, many country-year observations are missing, to minimize the impact of missing-not-at-random (countries with missing data systematically tend to be lower-income nations with weaker research infrastructure), we adopted two missingness filling methods to achieve higher coverage of country-level indicators.

1. Within-country longitudinal year-filling: For countries with observed indicator data in a subset of years, temporal gaps within each country series were resolved using domain-specific longitudinal gap-filling:

   Continuous socio-economic & health metrics (GDP per capita, Health Expenditure, HDI, Physician Density, Hospital Bed Density, R&D Expenditure, Researchers Density, Total Publications, Total Citations, UHC Index): Missing intermediate years between two observed data points were linearly interpolated. Missing years outside the range of observed values were filled using Last Observation Carried Forward (LOCF) or Next Observation Carried Backward (NOCB).

   Survey waves & slow-moving institutional indicators (Electoral Democracy Index, Trust in Government, Trust in Scientists, Cultural Values, World Giving Altruism Index, World Bank Income Group): Survey wave data were propagated forward (LOCF) and backward (NOCB) longitudinally within each country to maintain the country's institutional and cultural baseline.

   Static infrastructure indicators (Medical Schools): Total medical school counts (recorded for 2024) were assigned statically across 2000-2024 for each country.

   Impact of temporal filling: This longitudinal gap-filling dramatically rescued data coverage across benchmark country-years (e.g., Physician Density coverage increased from 58.9% to 95.6%; Hospital Bed Density from 61.3% to 94.5%; HDI from 89.2% to 95.1%; R&D Expenditure from 44.7% to 78.7%; and the UHC Index achieved 95.6% coverage across 175 countries).

2. Cross-country imputation (geographic subregion × income tier interaction imputation)

   For countries completely lacking data for specific indicators across all calendar years, missing values were imputed using a geographic subregion × World Bank Income Group interaction imputation. Countries were grouped into peer buckets defined by the interaction of geographic subregion and income classification tier (e.g., Sub-Saharan Africa × Low Income, Latin America & Caribbean × Upper-Middle Income). Missing values were imputed as the mean of peer countries within the same geographic and income tier. This interaction structure ensures that imputed values accurately reflect regional and economic peer benchmarks without distorting cross-country structural variance.

All data sources, country coverage rates (ranging from 57.4% to 100%), variable transformations, and inclusion decisions for the 20 indicators are reported in **Supplementary Table 7**.

## 5. Intervention Analysis Framework

### 5.1 Network evolution modeling

We estimated how network metrics change with reductions in structural inequality ($\Delta G$) by modeling dynamic network reconfiguration as structural misalignments are corrected. Under policy alignment, systemic barriers are dismantled, expanding co-participation across previously isolated country-factor pairs and promoting global network integration:

Density Expansion (D): Network density increases as co-participation expands:

$D_{new}=\min(0.95, D_{base}+D_{base}\cdot\Delta G\cdot 0.5\cdot E_{strat})$

Homophily Reduction ($H_{factor}$): Factor segregation decreases as structural misalignments are corrected, fostering cross-factor integration: $H_{new}=\max(0.10, H_{base} - H_{base}\cdot\Delta G\cdot 0.4\cdot E_{strat})$

Modularity Dissolution (Q): Community segregation decreases ($Q_{new}=\max(0, Q_{base} - Q_{base}\cdot\Delta G\cdot 0.3\cdot E_{strat})$), indicating broader inter-group collaboration across network clusters.

### 5.2 Intervention simulation algorithm:

For each intervention step (corresponding to Gini reduction $\Delta G=(G_{baseline}-G_{step})/G_{baseline}$):

1. Calculate new network metrics: Compute $D_{new}$, $H_{new}$, and $Q_{new}$ using the strategy-specific efficiency scaling multiplier ($E_{strat}=1.0$ for Full Alignment; $E_{strat}=1.6$ for Targeted Alignment, reflecting resource-efficient prioritization).
2. Apply factors: Enforce natural domain bounds, $0\leq\text{Density}\leq 0.95$, $0.10\leq\text{Homophily}\leq 1.00$, $\text{Modularity}\geq 0.00$.
3. Uncertainty & Measurement Bounds: Apply uncertainty bounds proportional to baseline variation (10% uncertainty at Step 0 baseline, scaling to 15% at final alignment step).
4. Efficiency Metrics: Intervention effects were estimated using bootstrap resampling across 200 iterations. Strategy efficiency ratios range from 1.46 × to 2.22× higher reduction per country adjusted for Targeted Alignment compared to Full Alignment.

### 5.3 Statistical testing and uncertainty estimation

Bootstrap Confidence Intervals: All intervention effects and network metric changes include 95% confidence intervals calculated from: 200 bootstrap samples for intervention effects and 100 parametric bootstrap samples for network metrics evolution.

Statistical significance tests:

- For testing our primary hypothesis tests, disease vs. country variance components, we did not apply multiple testing correction, as a single primary null hypothesis was tested.

- Gini reduction significance: For each intervention scenario, we tested whether post-intervention Gini distributions differed from baseline using paired t-tests across bootstrap samples. Both scenarios showed significant reductions ($p<0.001$).
- Strategy comparison: We compared full vs targeted alignment using: Difference in mean Gini reduction with 95% CI; paired t-test on bootstrap samples (t=-1030.76, p = 2.95e-836), mean Gini difference (–0.3181, 95% CI: [–0.3307, –0.3050]); Efficiency ratio with bootstrap confidence interval.
- Network metric changes: We tested whether final network metrics differed significantly from baseline using permutation tests (10,000 permutations).

# 6. Representativeness Analysis

## 6.1 Overview

We conducted a comprehensive representativeness analysis of RCT datasets across several filtering, extraction, and subsetting procedures to evaluate whether data refinement steps introduced systematic selection bias. To ensure that our final datasets faithfully reflect the broader population of clinical research, we benchmarked each subset against its parent baseline dataset across four factors that account for potential selection bias: Temporal distribution, geographic distribution, journal and disciplinary coverage, and topic distribution.

## 6.2 Data structure

The data refinement pipeline comprises five levels:

- 301k (FullRCT) [n = 301,262; 1980-2024]: The complete corpus of RCTs indexed in PubMed.
- 195k (TotalRCT) [n = 193,806; 2000-2024]: The temporally restricted corpus based on available geographic and institutional metadata.
- 137k (DisTSub) [n = 137,446; 2000-2024]: The clinical disease subset filtered to studies with matched Global Burden of Disease (GBD) cause-level classifications.
- 124k (GeoTSub) [n = 124,189; 2000-2024]: The updated dataset including geographic participation metadata, extracted via LLM and ClinicalTrials.gov registry mapping across 183 GBD-matched nations.
- 78k (DisGeoSub) [n = 78,117; 2000-2024]: The complete subset of RCTs with complete geographic, enrolling participants, and disease/cause metadata.

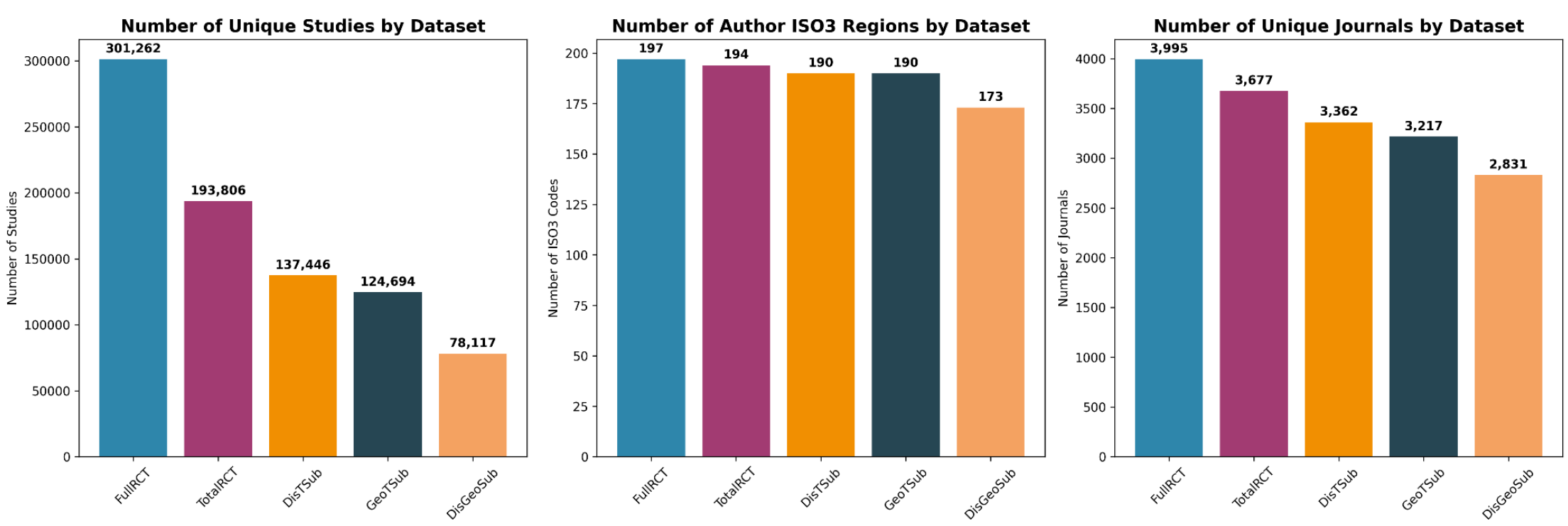

## 6.3 Effect Size

Traditional statistical hypothesis tests (e.g., Pearson's $\chi 2$ test) are sensitive to sample size. In very large datasets (n>100,000), negligible, even infinitesimal, distributional shifts yield significant results. To provide a scale-invariant assessment of the magnitude of the differences in the subset's representativeness, we calculated Cramér's V effect size. Values were interpreted as: <0.1 (very small: highly representative), 0.1-0.3 (small: representative), 0.3-0.5 (medium: moderately representative), and >0.5 (large: not representative).

## 6.4 Results of the representativeness analysis

Across all four factors (temporal distribution, geographic distribution, journal and disciplinary coverage, and topic distribution) and 16 individual comparisons, Cramér's V effect sizes remained consistently below the 0.10 threshold (V<0.10), confirming that the procedure followed to construct our dataset did not introduce selection bias at any step.

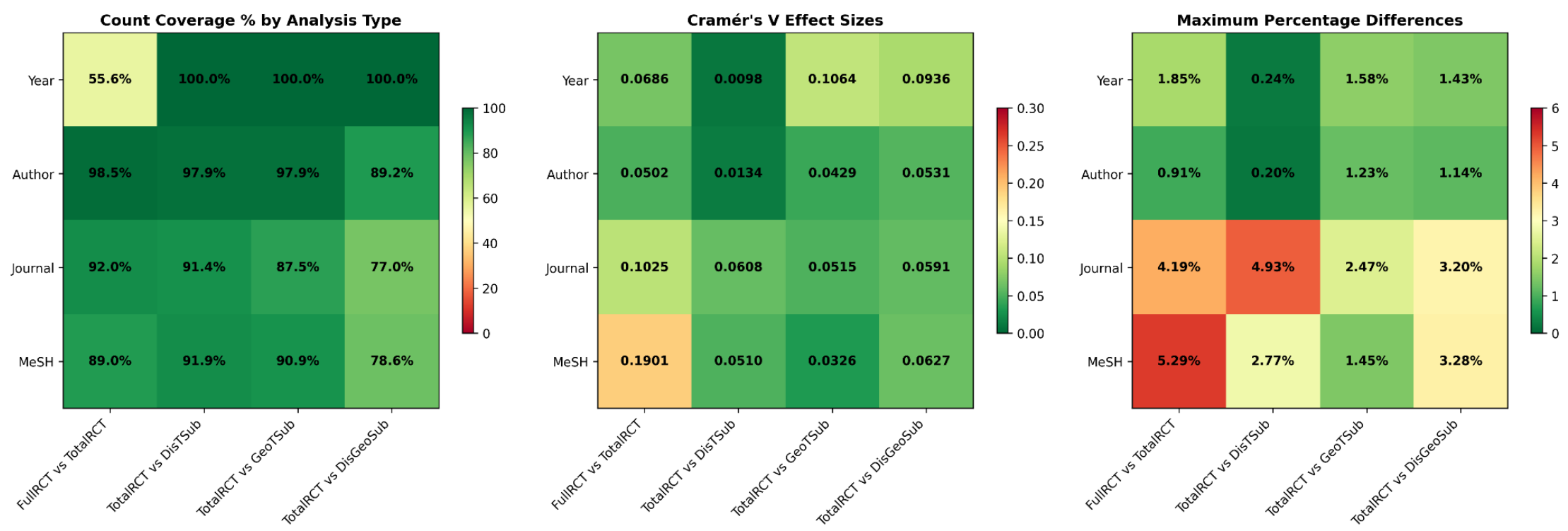

1. Temporal distribution (publication years)

- Year coverage: 100.0% matching between the four sub-datasets.
- Effect sizes:
    - FullRCT (301k) vs TotalRCT (195k): V=0.0686 (very small)
    - TotalRCT (195k) vs DisTSub (137k): V=0.0098 (very small)
    - TotalRCT (195k) vs GeoTSub (124k): V=0.01064 (small)
    - TotalRCT (195k) vs DisGeoSub (78k): V=0.0936 (very small)
- Maximum percentage differences per year ranged from 0.24% to 1.85%, with mean differences consistently below 0.62%.

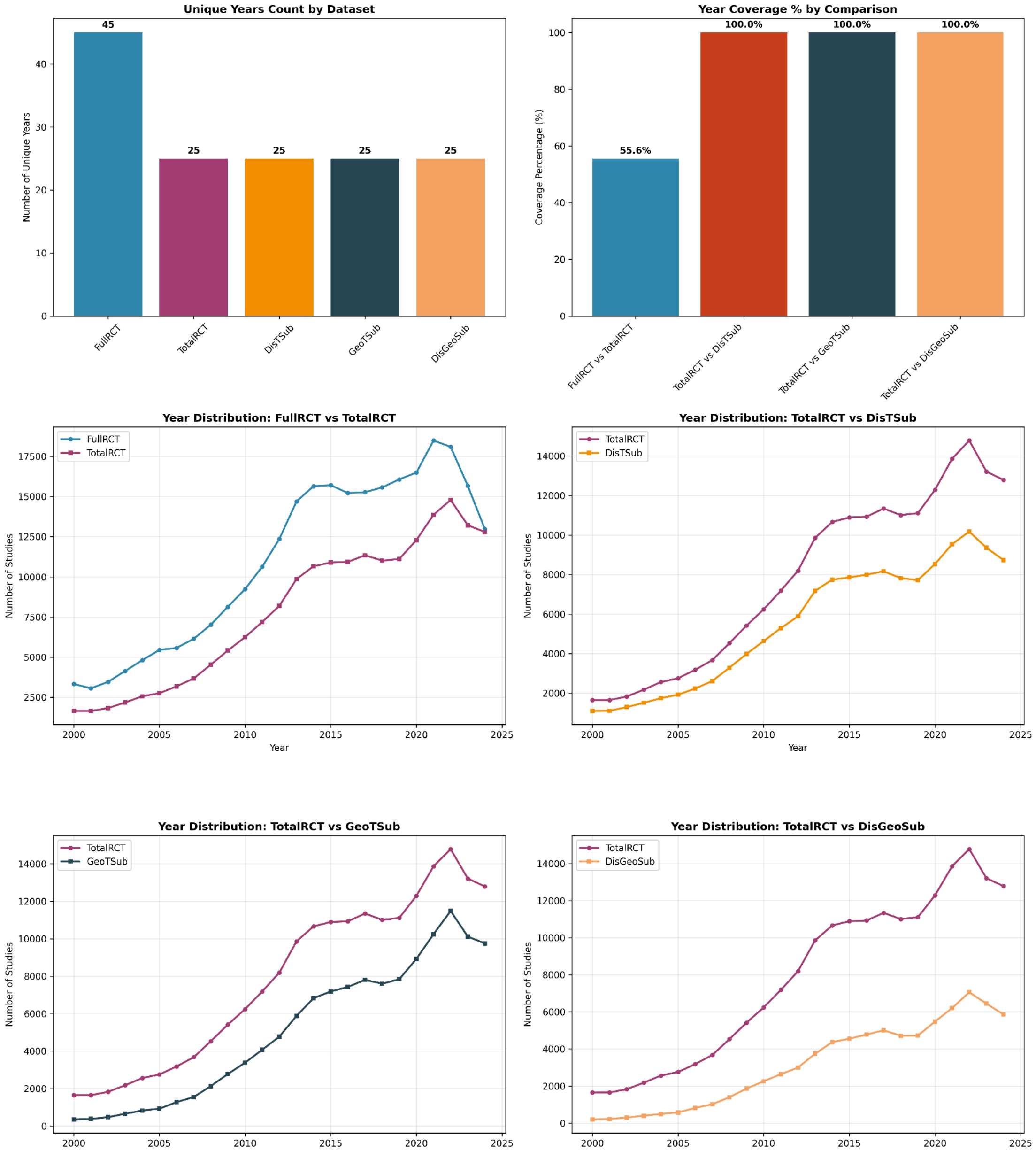

2. Geographic distribution (author affiliation countries)

- Country coverage: 89.2% to 98.5% ISO3 coverage, covering 173–194 region codes.
- Effect sizes:
    - FullRCT (301k) vs TotalRCT (195k): V=0.0502 (very small)
    - TotalRCT (195k) vs DisTSub (137k): V=0.0134 (very small)
    - TotalRCT (195k) vs GeoTSub (124k): V=0.0429 (very small)
    - TotalRCT (195k) vs DisGeoSub (78k): V=0.0531 (very small)
- Maximum country-level percentage differences were modest (0.20%-1.23%), with mean percentage differences below 0.04%.

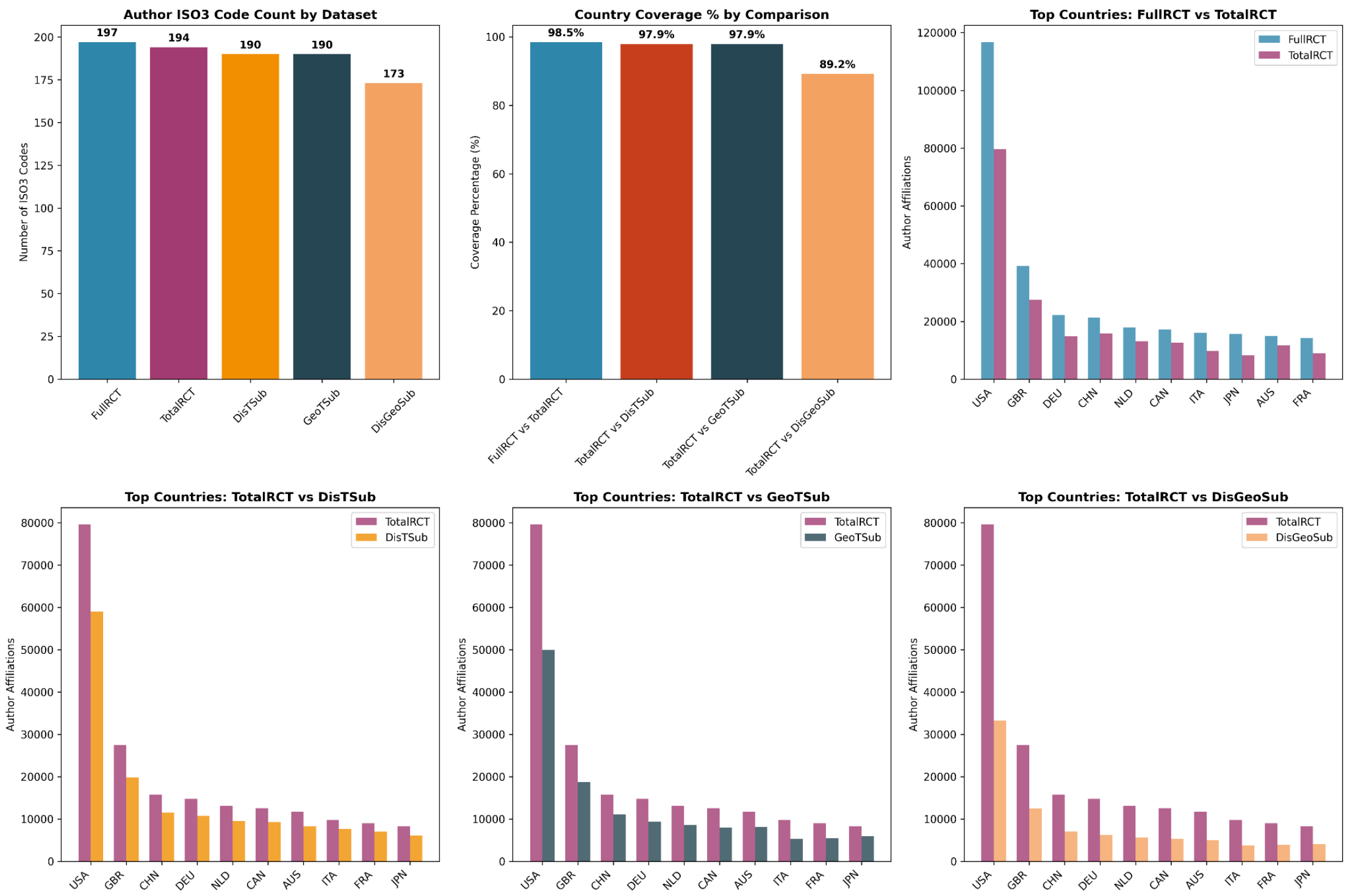

3. Journal and disciplinary coverage

- Journal: 77.0% to 92.0% unique journal retention; 100% coverage across 15 high-level NLM journal categories.
- Effect sizes:
  - FullRCT (301k) vs TotalRCT (195k): V=0.1025 (small)
  - TotalRCT (195k) vs DisTSub (137k): V=0.0608 (very small)
  - TotalRCT (195k) vs GeoTSub (124k): V=0.0515 (very small)
  - TotalRCT (195k) vs DisGeoSub (78k): V=0.0591 (very small)
- Maximum percentage differences in journal category distributions remained well within acceptable limits (2.31%-4.93%).

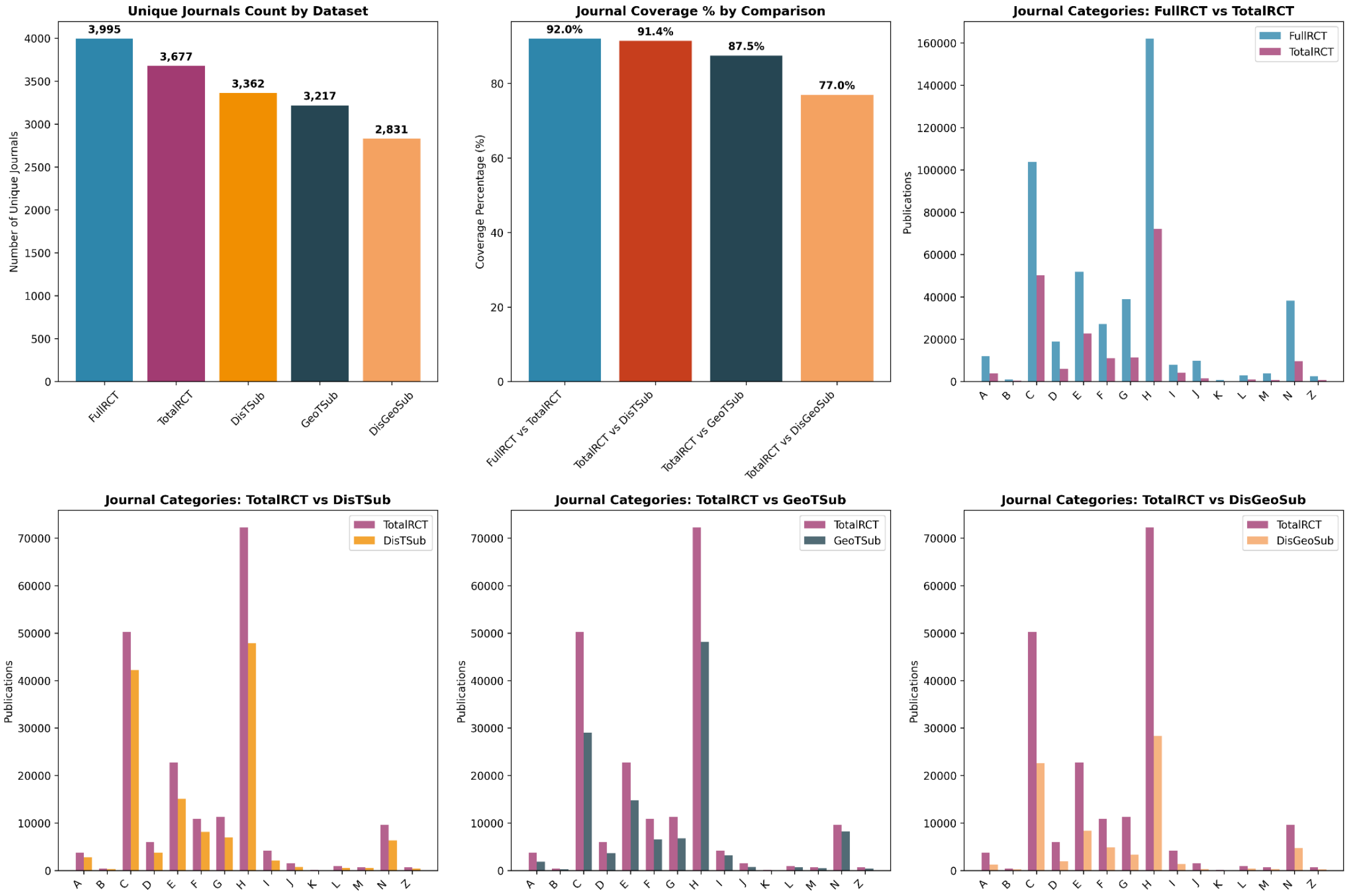

4. Topic distribution (MeSH, top level categories)

- Category Coverage: 78.6% to 91.9% unique MeSH term retention across 31,659-45,239 terms.
- Effect Sizes:
  - FullRCT (301k) vs TotalRCT (195k): V=0.1901 (small)
  - TotalRCT (195k) vs DisTSub (137k): V=0.0510 (very small)
  - TotalRCT (195k) vs GeoTSub (124k): V=0.0326 (very small)
  - TotalRCT (195k) vs DisGeoSub (78k): V=0.0627 (very small)
- Maximum percentage differences in research topic distributions ranged from 1.45% to 5.29%, with mean differences below 0.95%.

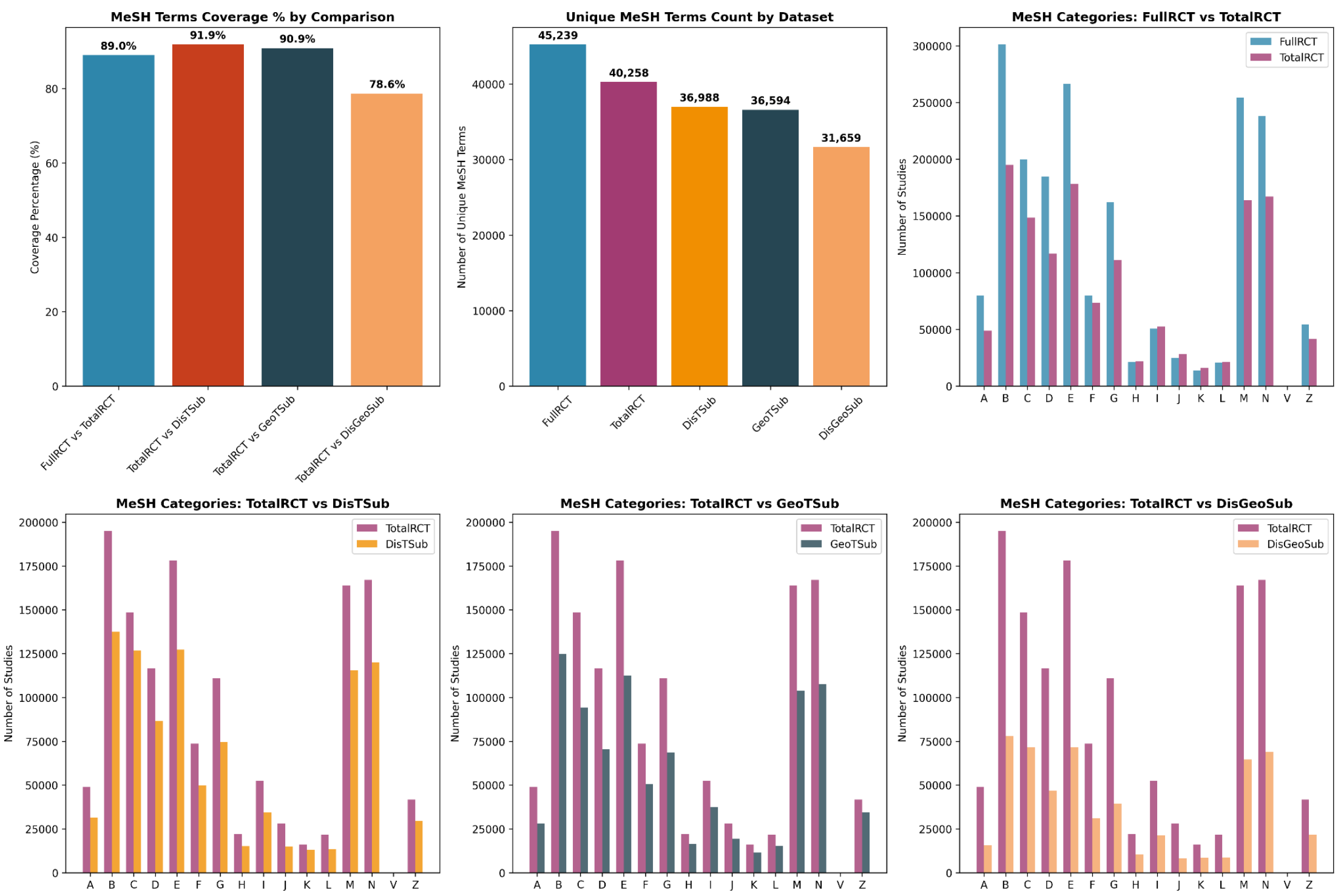

In all, the findings confirm that the filtering procedure from the original 301,262 RCTs down to the 124,189 geographic cohort and 78,117 disease-matched final datasets did not introduce a selection bias. Cramér's V effect sizes are well below the 0.10 benchmark for representativeness concern for each factor (temporal distribution, geographic distribution, journal and disciplinary coverage, and topic). Thus, the refined 124k and 78k datasets are representative of the global RCTs literature.